# Optical properties of current carrying molecular wires

Michael Galperin[†] and Abraham Nitzan

School of Chemistry, Tel Aviv University, Tel Aviv 69978, Israel

## Abstract

We consider several fundamental optical phenomena involving single molecules in biased metal-molecule-metal junctions. The molecule is represented by its highest occupied and lowest unoccupied molecular orbitals, and the analysis involves the simultaneous consideration of three coupled fluxes: the electronic current through the molecule, energy flow between the molecule and electron-hole excitations in the leads and the incident and/or emitted photon flux. Using a unified theoretical approach based on the non-equilibrium Green function method we derive expressions for the absorption lineshape (not an observable but a useful reference for considering yields of other optical processes) and for the current induced molecular emission in such junctions. We also consider conditions under which resonance radiation can induce electronic current in an unbiased junction. We find that current driven molecular emission and resonant light induced electronic currents in single molecule junctions can be of observable magnitude under appropriate realizable conditions. In particular, light induced current should be observed in junctions involving molecular bridges that are characterized by strong charge transfer optical transitions. For observing current induced molecular emission we find that in addition to the familiar need to control the damping of molecular excitations into the metal substrate the phenomenon is also sensitive to the way in which the potential bias is distributed on the junction.

[†] Present address: Department of Chemistry, Northwestern University, Evanston IL 60208, USA.

# 1. Introduction

The last few years have seen a surge in activity in studies of transport through molecular wires. Experimental techniques for putting and electrically monitoring single molecules in small gaps between metal leads range from lithography and deposition (for junctions involving carbon nanotubes) scanning probe spectroscopy (SPS; including scanning tunneling microscopy, STM, or conducting atomic force microscopy, AFM), sometimes aided by a gold nanoparticle as a directing device, break junction techniques, electromigration methods, and more. For recent reviews see [1]). A particularly interesting demonstration of directed assembly using an AFM tip was recently published[2]). A parallel effort exists in fabrication and electrical studies of molecular layers between metal leads.[3] Studies of the current voltage characteristics of the so-obtained junctions reveal many interesting phenomena such as non-ohmic response, rectification, negative differential resistance and switching. In addition, extensive studies of inelastic effects were carried out.[4] Most of the structures studied to date are two terminal junctions, but an effective gate potential could be achieved in a few cases either by using the substrate as a gate electrode[5] or by controlling the redox potential of an electrolyte environment.[6] Controlling the junction operation by structure manipulation has also been considered.[7] The field still faces formidable challenges of quantitative reproducibility (in particular between results obtained by different groups) and (probably a related issue) junction stability, however it is evident that single molecule operation exhibiting these phenomena seems by now to have been successfully demonstrated.

In addition to gating and structural control, the use of an external field as a controlling tool provides an obvious possibility,[8,9] however its application in the small nanogap between two metal leads is difficult to implement. Such effects were observed in larger mesoscopic structures,[10] and optical control of an electron transfer reaction in solution has been demonstrated.[11]. Recently, light induced switching behavior in the conduction properties of molecular nano-junctions has been demonstrated[12] and voltage effects on the fluorescence yield of molecules in such junctions were observed.[13] In addition to controlling transport with external radiation, other optical phenomena involving molecular junctions are of interest. For example, radiative and non-radiative lifetimes of excited molecules near metal surfaces have been observed and discussed[14] and molecular fluorescence induced by inelastic electron tunneling has been seen.[15] A recent observation of emission that accompanies electronic conduction through a silver nanodot,[16] enhanced in the presence of microwave radiation, was attributed to a nano-light emitting diode (LED) phenomenon. Finally, recent observation of "giant" surface enhanced Raman scattering (SERS)[17] was suggested to be associated with molecules positioned in narrow gaps between metal particles – another type of nanojunction. SERS[18] is attributed mainly to the local enhancement of the radiation field at rough features on certain noble metal surfaces, [19] [20,21] and related effects on other optical properties of molecules adsorbed on metal and dielectric surfaces have been discussed.[19] Additional contributions to the enhancement arise from first layer effects associated with electron sharing between molecular and metal orbitals,[22] and it has been suggested that electron motion through the molecule in metal-molecule-metal contacts will reduce the EM field enhancement and at the same time may open a new channel for Raman scattering.[23]

If experimental setups that can couple biased molecular wires to the radiation field could be achieved, general questions concerning the optical response of molecules in non-equilibrium situations come to mind. A general theory of the optical response of a molecule open to electron reservoirs *under bias* and during current flow is presently not available. It should be emphasized that while this issue is interesting as a fundamental questions, observation of optical phenomena in present experimental setups is not easy both because it is hard to inject light into molecular size slits between two metal leads and because a natural probe in such experiments – the molecular emission – may be strongly damped because of proximity to a metal surface. Nevertheless, in view of the observations already made and of the general potential importance of what may be termed "nanojunction spectroscopy" it is important to consider the properties of such systems. In this paper we begin to undertake this task by considering several fundamental optical phenomena involving single molecules in biased metal-molecule-metal junctions. Following the introduction of our model and theoretical methodology in Section 2, we address in Section 3 the issue of molecular light absorption in a biased and current carrying junction (probably not an observable but a useful input for estimating yields of other optical processes). In Section 4 we consider condition under which resonance radiation can induce electronic current in an unbiased junction and in Section 5 we study current induced light emission in molecular junctions. Section 6 concludes.

## 2. Model and method

We consider a molecule represented by its highest occupied molecular orbital (HOMO), $|1\rangle$, and lowest unoccupied molecular orbital (LUMO), $|2\rangle$, positioned between two leads represented by free electron reservoirs *L* and *R* and interacting with the radiation field (Fig 1). In the independent electron picture a transition between the ground and excited molecular states corresponds to transfer of an electron between levels $|1\rangle$ and $|2\rangle$. The reservoirs are characterized by the electronic chemical potentials $\mu_L$ and $\mu_R$, where the difference $\mu_L - \mu_R = e\Phi$ is the imposed voltage bias. In this picture the Hamiltonian is

$$\hat{H} = \hat{H}_0 + \hat{V} \tag{1}$$

$$\hat{H}_0 = \sum_{m=1,2} \varepsilon_m \hat{c}_m^\dagger \hat{c}_m + \sum_{k \in \{L,R\}} \varepsilon_k \hat{c}_k^\dagger \hat{c}_k + \hbar \sum_\alpha \omega_\alpha \hat{a}_\alpha^\dagger \hat{a}_\alpha \tag{2}$$

$$\hat{V} = \hat{V}_M + \hat{V}_P + \hat{V}_N \tag{3}$$

$$\hat{V}_M = \sum_{K=L,R} \sum_{m=1,2; k \in K} \left( V_{km}^{(MK)} \hat{c}_k^\dagger \hat{c}_m + h.c. \right) \tag{4}$$

$$\hat{V}_P = \left( V_0^{(P)} \hat{a}_0 \hat{c}_2^\dagger \hat{c}_1 + h.c \right) + \sum_{\alpha \neq 0} \left( V_\alpha^{(P)} \hat{a}_\alpha \hat{c}_2^\dagger \hat{c}_1 + h.c \right) \tag{5}$$

$$\hat{V}_N = \sum_{K=L,R} \sum_{k \neq k' \in K} \left( V_{kk'}^{(NK)} \hat{c}_k^\dagger \hat{c}_{k'} \hat{c}_2^\dagger \hat{c}_1 + h.c. \right) \tag{6}$$

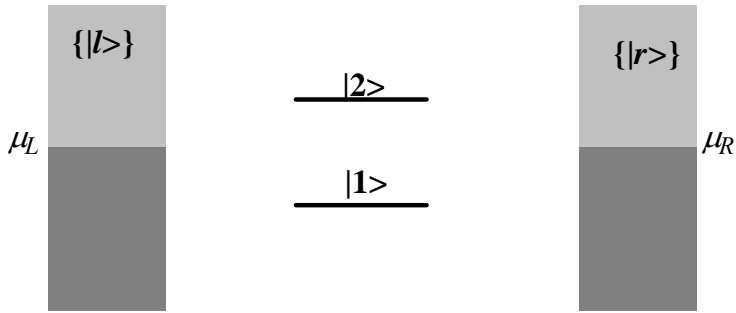

Fig.1. A model for light induced effects in molecular conduction. The right ($R=|\{r\}\rangle$) and left ($L=|\{l\}\rangle$) manifolds represent the two metal leads characterized by electrochemical potentials $\mu_R$ and $\mu_L$ respectively. The molecule is represented by its highest occupied molecular orbital (HOMO), $|1\rangle$, and lowest unoccupied molecular orbital (LUMO), $|2\rangle$.

where $L$ and $R$ denote the left and right leads, respectively, and h.c. denotes hermitian conjugate. The Hamiltonian $\hat{H}_0$, Eq. (2), contains additively terms that correspond to the isolated molecule, the free leads and the radiation field while Eqs. (3)-(6) describe the coupling between these subsystems. Here the operators $\hat{c}$ and $\hat{c}^\dagger$ are annihilation and creation operators of an electron in the various states, while $\hat{a}$ and $\hat{a}^\dagger$ are the corresponding operators for photons. $\hat{V}_M$ and $\hat{V}_N$ denote two types of couplings between the molecule and the metal leads: $\hat{V}_M$ describes electron transfer that gives rise to net current in the biased junction, while $\hat{V}_N$ describes energy transfer between the molecule and electron-hole excitations in the leads. The latter is written in the near field approximation, disregarding retardation effects that will be important at large molecule-lead distances.[14]

$\hat{V}_P$, Eq. (5), is the molecule-radiation field coupling. Since we consider driving of the junction by an electromagnetic field as well as current induced spontaneous emission from the junction, we have written explicitly the term that corresponds to mode "0" that pumps the system. With regards to optical processes, we limit ourselves to near resonance processes pertaining to linear spectroscopy. This justifies the use of the rotating wave approximation (RWA) in Eq. (5). Also for this reason we may consider only zero- and one-photon states of the radiation field, take $V_0^{(P)}$ (but not $V_{\alpha\neq 0}^{(P)}$) to be proportional to the incident field amplitude $\mathcal{E}_0$, and treat all processes to second order in this coupling. We note that all coefficients $V_\alpha^{(P)}$ reflect properties of the local electromagnetic field at the molecular bridge which depend in turn on the metallic boundary conditions. We will not address these issues explicitly in the present work but they will obviously be important in any detailed calculation involving interaction of molecular conduction junctions with the radiation field. In addition, the coefficients $V_\alpha^{(P)}$ depend on the photon frequency $\omega_\alpha$ because of the usual $\sqrt{\omega_\alpha}$ term in the bare molecule-radiation field coupling as well as from the $\omega_\alpha$ dependence of the reaction field that results from the frequency-dependent dielectric response of the metal leads. In what follows we disregard this dependence assuming that all relevant couplings can be evaluated at the molecular frequency $\omega_{21} = (\varepsilon_2 - \varepsilon_1)/\hbar$.

In the Keldysh non-equilibrium Green's function (NEGF) formalism the steady state flux associated with a particular process $B$ is obtained from the system Green functions and the associated self energies by

$$I_B = \int_{-\infty}^{\infty} \frac{dE}{2\pi\hbar} \text{Tr}\left[\Sigma_B^<(E) G^>(E) - \Sigma_B^>(E) G^<(E)\right] \tag{7}$$

where, as above, all functions are defined in the "system" (molecular bridge) subspace. In (7) $\Sigma_B^<$ and $\Sigma_B^>$ are the self energies associated with the process $B$, and the trace is over the system states. Note that Eq. (7) was first derived[24] for electron current through a junction connecting two leads. The same formalism can be applied to other electronic fluxes, e.g. fluxes between bridge orbitals by replacing the originally considered hopping of electrons between lead and bridge by terms associated with in/out scattering of electrons between these molecular orbitals. Also, because in linear spectroscopy, optically induced transitions between ground and excited molecular states are accompanied by photon absorption and emission, the electronic flux associated with this transition accounts also for the corresponding photon absorption/emission flux, that is, describes the absorption (emission) lineshape.

Consider first the model without the radiative coupling $\hat{V}_P$ and the non-radiative energy transfer $\hat{V}_N$, i.e.

$$\hat{\tilde{H}}_0 = \hat{H}_0 + \hat{V}_M \tag{8}$$

This model contains only one particle operators and is exactly soluble. In the wide band approximation the retarded and advanced self-energies are taken purely imaginary and energy independent. We also take them to be diagonal in the $1-2$ representation

$$\Sigma^r_{MK} = \left(\Sigma^a_{MK}\right)^* = \begin{pmatrix} -i\Gamma_{MK,1}/2 & 0 \\ 0 & -i\Gamma_{MK,2}/2 \end{pmatrix} \tag{9}$$

where $K = L, R$ denotes the left and right electrode, respectively. Consequently, the retarded and advanced GFs in the molecular subspace $(1,2)$ are given by

$$G^r(E) = \begin{pmatrix} \dfrac{1}{E-\varepsilon_1 + i\Gamma_{M,1}/2} & 0 \\ 0 & \dfrac{1}{E-\varepsilon_2 + i\Gamma_{M,2}/2} \end{pmatrix}; \quad G^a(E) = \left[G^r(E)\right]^* \tag{10}$$

where $\Gamma_{M,m} = \Gamma_{ML,m} + \Gamma_{MR,m}$ with $m = 1,2$. In the same approximation the lesser and greater SEs are given by

$$\Sigma_M^{>,<} = \Sigma_{ML}^{>,<} + \Sigma_{MR}^{>,<} \tag{11a}$$

$$\Sigma_{MK}^{<}(E) = \begin{pmatrix} if_K(E)\Gamma_{MK,1} & 0 \\ 0 & if_K(E)\Gamma_{MK,2} \end{pmatrix} \tag{11b}$$

$$\Sigma_{MK}^{>}(E) = \begin{pmatrix} -i[1-f_K(E)]\Gamma_{MK,1} & 0 \\ 0 & -i[1-f_K(E)]\Gamma_{MK,2} \end{pmatrix} \tag{11c}$$

The lesser and greater GFs in the molecular subspace can then be obtained from the Keldysh formula

$$G^{<,>}(E) = G^r(E)\,\Sigma^{<,>}(E)\,G^a(E) \tag{12}$$

In these expressions

$$\Gamma_{MK,m} = 2\pi \sum_{k \in K} \left|V_{km}^{(MK)}\right|^2 \delta(E-\varepsilon_k),\ m=1,2\ \text{and}\ K=L,R \tag{13}$$

and $f_K(E)$, $K=L,R$, are the Fermi functions

$$f_K(E) = \left[\exp((E-\mu_K)/k_B T)+1\right]^{-1} \tag{14}$$

Eqs. (9)-(12) lead to the well known Landauer formula for electrical conduction that yields the source-drain electronic current $I_{sd}$. For our model the result is obtained as a sum over currents through the ground and excited molecular levels (hole and electron currents, respectively)

$$I_{sd} = \frac{1}{\hbar}\int_{-\infty}^{+\infty}\frac{dE}{2\pi}\sum_{m=1,2}\Gamma_{ML,m}\,G_{mm}^r(E)\,\Gamma_{MR,m}\,G_{mm}^a(E)[f_L(E)-f_R(E)] \tag{15}$$

In the presence of the radiative and non-radiative energy transfer couplings, $\hat{V}_P$ and $\hat{V}_N$, four fluxes come into balance at steady state: The absorbed and emitted photon fluxes, $I_{abs}$ and $I_{em}$, the non radiative relaxation $I_N$ and the electrical current $I_{sd}$. To describe this situation we now consider the full Hamiltonian (1)-(6). To calculate the needed self energies we treat the perturbations $\hat{V}$ in the standard lowest nonzero (second) order in interaction on the Keldysh contour [25] and use the non-crossing approximation (NCA)[26] whereupon a self-energy associated with a given process is taken to be decoupled from interactions associated with other processes. The total SE is then given by a sum of contributions associated with different processes

$$\Sigma = \Sigma_{ML} + \Sigma_{MR} + \Sigma_P + \Sigma_{NL} + \Sigma_{NR} \qquad (16)$$

On the Keldysh contour these self energies are (to second order; see Appendix A)

$$\Sigma_{MK}(\tau_1,\tau_2) = \begin{bmatrix} \Sigma_{MK,11}(\tau_1,\tau_2) & \Sigma_{MK,12}(\tau_1,\tau_2) \\ \Sigma_{MK,21}(\tau_1,\tau_2) & \Sigma_{MK,22}(\tau_1,\tau_2) \end{bmatrix} \qquad (17a)$$

$$\Sigma_{MK,mm'}(\tau_1,\tau_2) = \sum_{k \in K} V_{mk}^{(MK)} g_k(\tau_1,\tau_2) V_{km'}^{(MK)} \qquad (17b)$$

$$\Sigma_P(\tau_1,\tau_2) = i \sum_\alpha \left|V_\alpha^{(P)}\right|^2 \begin{bmatrix} F_\alpha(\tau_2,\tau_1)G_{22}(\tau_1,\tau_2) & 0 \\ 0 & F_\alpha(\tau_1,\tau_2)G_{11}(\tau_1,\tau_2) \end{bmatrix} \qquad (18)$$

$$\Sigma_{NK}(\tau_1,\tau_2) = \sum_{k \neq k' \in K} \left|V_{kk'}^{(NK)}\right|^2 g_k(\tau_2,\tau_1) g_{k'}(\tau_1,\tau_2) \begin{bmatrix} G_{22}(\tau_1,\tau_2) & 0 \\ 0 & G_{11}(\tau_1,\tau_2) \end{bmatrix} \qquad (19)$$

where again $K=L, R$ and where $g_k$ and $F_\alpha$ are free electron GFs in state $k$ and free photon GFs of the mode α, respectively.[1]

After projection onto the real time axis we get the retarded, advanced, lesser, and greater components of these SEs, which, in steady state situations can be expressed in energy space. The SEs associated with electron exchange between molecule and leads,

---

[1] $F_\alpha(\tau_1,\tau_2) = -i\langle T_c\, \hat{a}_\alpha(\tau_1)\hat{a}_\alpha^\dagger(\tau_2)\rangle$ and $g_k(\tau_1,\tau_2) = -i\langle T_c \hat{c}_k(\tau_1) c_k^\dagger(\tau_2)\rangle$ where $T_c$ is the contour ordering operator. Note that if we do not make the rotating wave approximation (RWA) the $\hat{a}_\alpha + \hat{a}_\alpha^\dagger$ would replace $\hat{a}_\alpha$ and $\hat{a}_\alpha^\dagger$ everywhere and the corresponding photon GF would be
$D_\alpha(\tau_1,\tau_2) = -i\langle T_c(\hat{a}_\alpha(\tau_1) + \hat{a}_\alpha^\dagger(\tau_1))(\hat{a}_\alpha(\tau_2) + \hat{a}_\alpha^\dagger(\tau_2))\rangle$

$\Sigma_{MK}$, were already given in Eqs. (9) and (11) for a model that assumes that no interstate mixing results from coupling to the metals. The lesser and greater SEs associated with the radiative coupling are easily obtained by applying the Langreth relations[27] to Eq. (18). We get

$$\Sigma_P^<(E) = \sum_\alpha |V_\alpha^{(P)}|^2 \begin{bmatrix} (1+N_\alpha)G_{22}^<(E+\omega_\alpha) & 0 \\ 0 & N_\alpha G_{11}^<(E-\omega_\alpha) \end{bmatrix} \quad (20a)$$

$$\Sigma_P^>(E) = \sum_\alpha |V_\alpha^{(P)}|^2 \begin{bmatrix} N_\alpha G_{22}^>(E+\omega_\alpha) & 0 \\ 0 & (1+N_\alpha)G_{11}^>(E-\omega_\alpha) \end{bmatrix} \quad (20b)$$

where $N_\alpha$ is the number of photons in mode $\alpha$. To obtain (20) we have used the GFs of a free photon field

$$F_\alpha^<(\omega) = -2\pi i N_\alpha \delta(\omega-\omega_\alpha) \quad F_\alpha^>(\omega) = -2\pi i (1+N_\alpha)\delta(\omega-\omega_\alpha) \quad (21)$$

As will be seen below, the sum over α can be restricted to modes of interest. We need to include only the pumping mode, α = 0, for the calculation of the absorption flux, only modes of a given frequency (within the resolution window) to get the frequency resolved emission and all modes (no restrictions) in order to compute the total emission flux.

Finally, to get the greater and lesser SEs associated with energy transfer to electron-hole excitations in the leads we apply the Langreth rules to Eq.(19). This leads to

$$\Sigma_{NK}^<(E) = \int \frac{d\omega}{2\pi} B_{NK}(\omega,\mu_K) \begin{bmatrix} G_{22}^<(E+\omega) & 0 \\ 0 & G_{11}^<(E+\omega) \end{bmatrix} \quad (22a)$$

$$\Sigma_{NK}^>(E) = \int \frac{d\omega}{2\pi} B_{NK}(\omega,\mu_K) \begin{bmatrix} G_{22}^>(E-\omega) & 0 \\ 0 & G_{11}^>(E-\omega) \end{bmatrix} \quad (22b)$$

where $\mu_K$ is the chemical potential of the lead $K = L, R$ and

$$B_{NK}(\omega,\mu_K) = \int \frac{dE}{2\pi} C_{NK}(E,\omega) f_K(E)\left[1 - f_K(E+\omega)\right] \tag{23}$$

$$C_{NK}(E,\omega) = (2\pi)^2 \sum_{k \neq k' \in K} \left|V_{kk'}^{(NK)}\right|^2 \delta(E - \varepsilon_k) \delta(E + \omega - \varepsilon_{k'}) \tag{24}$$

In obtaining the expressions we have used the free electron lesser and greater GFs for the leads

$$g_k^<(E) = 2\pi i f_K(E) \delta(E - \varepsilon_k) \quad ; \quad g_k^>(E) = -2\pi i \left[1 - f_K(E)\right] \delta(E - \varepsilon_k) \tag{25}$$

The retarded and advanced SEs associated with these processes are more difficult to calculate from the Langreth rules. An alternative route using the Lehmann representation[28]

$$\Sigma^r(E) = i \int_{-\infty}^{+\infty} \frac{dE'}{2\pi} \frac{\Sigma^>(E') - \Sigma^<(E')}{E - E' + i\delta} \tag{26}$$

is also problematic because of the singularity in the integrand. One can circumvent the difficulty by assuming, in the spirit of the wide band approximation, that all diagonal components of $\Sigma$ are purely imaginary, in which case (26) yields for such components[2,3]

$$\Sigma^r(E) = \frac{1}{2}\left[\Sigma^>(E) - \Sigma^<(E)\right] \equiv -\frac{1}{2} i\Gamma \quad ; \quad \Sigma^a(E) \equiv \frac{1}{2} i\Gamma \tag{27}$$

Note that Eq.(27) is compatible with Eq.(9). Using this expression yields the retarded and advanced components of the SE (16) and the corresponding retarded and advanced GFs

---

[2] Eq.(27) may also be derived from the general equality $\Sigma^r(E) - \Sigma^a(E) = \Sigma^>(E) - \Sigma^<(E)$ under the same assumption

[3] While making the wide band approximation here is consistent with making it in similar contexts elsewhere, it should be remarked that assuming that an analytic casual function is purely imaginary is incompatible with Kramers-Kronig relationships. As with all approximations of this kind there is always an underlying assumption that the real part of $\Sigma$ is small and, anyway, was absorbed into the level energies implemented in $\hat{H}_0$.

$$G^r(E) = \begin{pmatrix} \dfrac{1}{E - \varepsilon_1 - \Sigma_{11}^r(E)} & 0 \\ 0 & \dfrac{1}{E - \varepsilon_2 - \Sigma_{22}^r(E)} \end{pmatrix} \quad (28)$$

In most cases the radiative contribution $\Sigma_{P,mm}^r$ $(m = 1,2)$ can be disregarded relative to the other width parameters and we neglect it in our calculations. The lesser and greater components of $\Sigma_P$ cannot be ignored however since they enter into the calculation of the radiative flux according to Eq.(7). For these flux calculations we also need the greater and lesser GFs that are obtained from the Keldysh equation (12).

With regard to the radiative fluxes considered in this work, we have distinguished between the absorption flux $I_{abs}$ and the spontaneous emission flux $I_{em}$. The former is associated with the pumping mode and is computed using Eq.(7) with the lesser and greater SEs associated with that mode

$$\Sigma_{P0}^<(E) = |V_0^{(P)}|^2 \begin{bmatrix} 2G_{22}^<(E + \omega_0) & 0 \\ 0 & G_{11}^<(E - \omega_0) \end{bmatrix} \quad (29a)$$

$$\Sigma_{P0}^>(E) = |V_0^{(P)}|^2 \begin{bmatrix} G_{22}^>(E + \omega_0) & 0 \\ 0 & 2G_{11}^>(E - \omega_0) \end{bmatrix} \quad (29b)$$

Eqs. (29) are obtained from the general expression (20) by considering only a single term $\alpha = 0$ with $N_0 = 1$. Regarding the spontaneous emission flux we may again consider the frequency resolved emission (differential emission flux) $I'_{em}(\omega) = dI_{em}(\omega)/d\omega$, and the total integrated emission $I_{em}^{tot} = \int_0^\infty d\omega I'_{em}(\omega)$. The differential (frequency resolved) flux $I'_{em}(\omega)$ is calculated from Eq. (7) using the self energy (20) with $N_\alpha = 0$ and with the sum over α restricted to modes of frequency ω. This leads to[4]

$$\Sigma_P^<(E,\omega) = \frac{\gamma_P(\omega)}{2\pi\rho_p(\omega)} \begin{bmatrix} G_{22}^<(E+\omega) & 0 \\ 0 & 0 \end{bmatrix} \quad (30a)$$

$$\Sigma_P^>(E,\omega) = \frac{\gamma_P(\omega)}{2\pi\rho_p(\omega)} \begin{bmatrix} 0 & 0 \\ 0 & G_{11}^>(E-\omega) \end{bmatrix} \quad (30b)$$

where

$$\gamma_P(\omega) = 2\pi \sum_\alpha |V_\alpha^{(P)}|^2 \delta(\omega - \omega_\alpha) = 2\pi \left(|V^{(P)}|^2\right)_\omega \rho_P(\omega) \quad (31)$$

and $\rho_P(\omega)$ is photon density of modes

$$\rho_P(\omega) = \frac{\omega^2}{\pi^2 c^3} \quad (32)$$

The frequency resolved flux is then obtained from Eq. (7) in the form[5]

$$I_{em}'(\omega) = \rho_p(\omega) \int_{-\infty}^{\infty} \frac{dE}{2\pi\hbar} \text{Tr}\left[\Sigma_p^<(E,\omega)G^>(E) - \Sigma_p^>(E,\omega)G^<(E)\right] \quad (33)$$

The self energy associated with the total emission flux is

$$\Sigma_p^<(E) = \int_0^\infty d\omega \rho(\omega) \Sigma_p^<(E,\omega) = \begin{pmatrix} \int_0^\infty \frac{d\omega}{2\pi} \gamma_P(\omega) G_{22}^<(E+\omega) & 0 \\ 0 & 0 \end{pmatrix} \quad (34a)$$

---

[4] Note that $\gamma_P(\omega)/(2\pi\rho_P(\omega)) = \left(|V^{(P)}|^2\right)_\omega$, where Eq. (31) may be regarded as the definition of $\left(|V^{(P)}|^2\right)_\omega$. Also note that the radiative width $\gamma_P$ and $\Gamma_P$ defined in accordance with Eq. (27), i.e. $\Gamma_P = i(\Sigma^> - \Sigma^<)$, are not the same (see, e.g. Eq. (38))

[5] Note that including ρ in Eqs. (30) and (33) did not make a difference to the final result, but is required to get the proper form and the correct dimensionality of the self energy $\Sigma_p^<(E,\omega)$ in Eq. (30)

$$\Sigma_P^>(E) = \int_0^\infty d\omega \rho(\omega) \Sigma_P^>(E,\omega) = \begin{pmatrix} 0 & 0 \\ 0 & \int_0^\infty \frac{d\omega}{2\pi} \gamma_P(\omega) G_{11}^>(E-\omega) \end{pmatrix} \quad (34b)$$

and the total emission flux is

$$I_{em}^{tot} = \int_{-\infty}^\infty \frac{dE}{2\pi\hbar} \text{Tr}\left[\Sigma_P^<(E)G^>(E) - \Sigma_P^>(E)G^<(E)\right] \quad (35)$$

To end this section we note that the calculations of the total emission flux, $I_{em}^{tot}$, and the flux associated with the non-radiative energy transfer to electron-hole excitations in the leads are relatively difficult because the evaluation of the relevant self energies requires integration over a frequency variable as seen in Eqs. (22) and (34). These calculations can be made simpler by using approximations for these self energies. In Appendix B we show that if $\varepsilon_{21}$ is large relative to the widths of levels 1 and 2 and provided some other modest assumptions are satisfied, then the following results provide good approximations for our applications

$$\Sigma_{NK}^< \simeq iB_{NK} \begin{pmatrix} n_2 & 0 \\ 0 & 0 \end{pmatrix} \quad (36a)$$

$$\Sigma_{NK}^> \simeq -iB_{NK} \begin{pmatrix} 0 & 0 \\ 0 & 1-n_1 \end{pmatrix} \quad (36b)$$

$$\Gamma_{NK} = i\left(\Sigma_{NK}^> - \Sigma_{NK}^<\right) = B_{NK} \begin{pmatrix} n_2 & 0 \\ 0 & 1-n_1 \end{pmatrix} \quad (37)$$

$$\Sigma_P^<(E) = i\gamma_P(\varepsilon_{21}) \begin{bmatrix} n_2 & 0 \\ 0 & 0 \end{bmatrix} \quad (38a)$$

$$\Sigma_P^>(E) = -i\gamma_P(\varepsilon_{21}) \begin{bmatrix} 0 & 0 \\ 0 & 1-n_1 \end{bmatrix} \quad (38b)$$

$$\Gamma_P = i\left(\Sigma_P^> - \Sigma_P^<\right) = \gamma_P(\varepsilon_{21})\begin{pmatrix} n_2 & 0 \\ 0 & 1-n_1 \end{pmatrix} \qquad (39)$$

In Eqs. (36)-(39) $n_1$ and $n_2$ are occupations of the bridge HOMO and LUMO states respectively.

In the following sections we use this procedure to study the behavior of three observables. In section 3 we evaluate the absorption line shape and the way it depends on the electrical driving. In section 4 we study the effect of the electromagnetic driving on the electronic current. Finally, in Section 5, we examine the fluorescence behavior of the driven molecule. Absorption spectrum is obviously not a likely observable for molecules embedded between two metal leads, however the effect of incident electromagnetic field on the molecular conduction behavior as well the possibility to drive fluorescence by external potential bias have been discussed and demonstrated in different contexts before. The present work provides a unified framework for describing and analyzing these phenomena.

## 3. Absorption line shape of a molecular bridge in a biased junction

As said above, the absorption line shape of a molecular bridge is not an easy observable. The results obtained below should therefore be regarded as an exemplary application of the general formulation of Section 2 rather than as theoretical predictions concerning realistic future experiments. The absorption flux can be calculated as the net flux induced by the pumping mode 0 through state $|2\rangle$

$$I_{abs}(\omega_0) = \int_{-\infty}^{\infty} \frac{dE}{2\pi\hbar}\left[\Sigma_{P0,22}^<(E)G_{22}^>(E) - \Sigma_{P0,22}^>(E)G_{22}^<(E)\right] \qquad (40)$$

or equivalently as the flux associated with that mode through state $|1\rangle$ with sign reversal

$$I_{abs}(\omega_0) = -\int_{-\infty}^{+\infty} \frac{dE}{2\pi\hbar} \left[ \Sigma^<_{P0,11}(E) G^>_{11}(E) - \Sigma^>_{P0,11}(E) G^<_{11}(E) \right] \quad (41)$$

The equality of the fluxes calculated from Eqs. (40) and (41) provides a convenient consistency check.

The SEs and GFs needed for the calculation of these fluxes are obtained by employing a self-consistent procedure starting from the standard metal-molecule-metal model defined by the Hamiltonian $\hat{\tilde{H}}_0 = \hat{H}_0 + \hat{V}_M$. In what follows we refer to this as our zero order description. The calculation proceeds as follows:

1. Calculate the zero-order GFs using (9) and (11) in (10) and (12).

2. Use these GFs in Eqs. (22), (29) and (30) to get a first iteration result for the SEs $\Sigma^{<,>}_{P0}$, $\Sigma^{<,>}_P$ and $\Sigma^{<,>}_{NK}$ $(K=L,R)$. Use Eq.(27) to get the corresponding results for $\Sigma^r$ and $\Sigma^a$. As already said (below Eq.(28)) the (spontaneous) radiative contribution $\Sigma^r_P$ to $\Sigma^r$ can be ignored. We also disregard the driving mode contribution $\Sigma^r_{P0}$ to $\Sigma^r$, since we are interested in the lowest order theory in the coupling to this mode (i.e. in the intensity of the incident field).

3. Use the calculated SEs in Eqs. (12) and (28) to update the GFs. This completes one iteration step.

4. The calculation proceeds by repeating steps 2 and 3 until convergence.

5. Convergence is declared when the populations $n_1$ and $n_2$ in the HOMO and LUMO states reach static values within a predefined tolerance (typically taken $10^{-6}$). These populations are computed from

$$in_m = \int \frac{dE}{2\pi} G^<_{mm}(E) \; ; \quad \text{or} \quad -i(1-n_m) = \int \frac{dE}{2\pi} G^>_{mm}(E) \; ; \quad (m=1,2) \quad (42)$$

6. After convergence is achieved, calculate the absorption flux using Eq.(40) or (41).

It should be noted that for the parameters used in this paper practical convergence is achieved already after the first iteration. Level population changes somewhat on subsequent iterations however the results shown in Fig.2 remain practically the same.

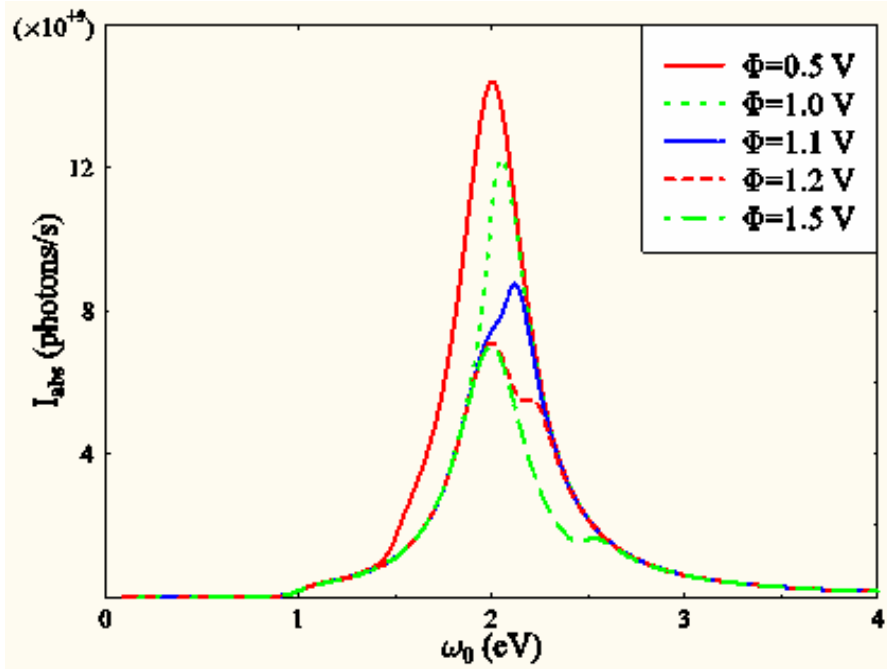

Fig.2 The absorption current (photons/s), Eq.(40) or (41) for the molecular model of Fig.1. The molecular electronic levels are assumed pinned to the right electrode, i.e. the bias shifts upward the electronic states of the left electrode. See text for parameters.

Fig.2 shows the results obtained from this calculation. We use the model of Fig.1 where, for the unbiased junction the metal Fermi levels are taken at mid point between levels $|1\rangle$ and $|2\rangle$. The parameters taken are $\varepsilon_{21} = 2\,eV$, $\gamma_P = 10^{-6}\,eV$, $B_{NL} = B_{NR} = 0.1\,eV$, $T = 300\,K$, $\Gamma_{ML,1} = \Gamma_{MR,1} = 0.01\,eV$ and $\Gamma_{ML,2} = \Gamma_{MR,2} = 0.2\,eV$. The potential bias is taken as a change of $\mu_L$, keeping the molecular energies pinned to the Fermi level of the right lead. Fig.2 depicts the line shape calculated as described above for several bias potentials. We note that the choice $\Gamma_{MK,1} \ll \Gamma_{MK,2}$ $(K = L, R)$ (implying the assumption that the HOMO is much more localized on the bridge than the LUMO) enhances the effect seen in Fig.2 – distortion of the line shape due to partial population of the LUMO. If the HOMO is broad as well this effect is smeared by integration over the HOMO density of states.

Some insight into these result can be obtained from an approximate analytical expression that can be derived by using the simplified forms of $\Sigma_P^{>,<}$ and $\Sigma_{NK}^{>,<}, K = L, R$ derived in Appendix B and keeping terms up to second order in the coupling to the pumping mode. This leads to (see Appendix C)

$$I_{abs} = \frac{|V_0^{(P)}|^2}{\hbar} \int_{-\infty}^{+\infty} \frac{dE}{2\pi} \frac{1}{(E-\varepsilon_2)^2 + (\Gamma_2/2)^2} \frac{1}{(E-\omega_0-\varepsilon_1)^2 + (\Gamma_1/2)^2} \times$$
$$\left\{ \left[ f_L(E-\omega_0)\Gamma_{ML,1} + f_R(E-\omega_0)\Gamma_{MR,1} + (B_N + \gamma_P)n_2 \right] \right. \qquad (43)$$
$$\times \left[ [1-f_L(E)]\Gamma_{ML,2} + [1-f_R(E)]\Gamma_{MR,2} + (B_N + \gamma_P)(1-n_1) \right]$$
$$\left. -2\left[ [1-f_L(E-\omega_0)]\Gamma_{ML,1} + [1-f_R(E-\omega_0)]\Gamma_{MR,1} \right]\left[ f_L(E)\Gamma_{ML,2} + f_R(E)\Gamma_{MR,2} \right] \right\}$$

where

$$\Gamma_m = \Gamma_{ML,m} + \Gamma_{MR,m} + \Gamma_{NL,m} + \Gamma_{NR,m} + \Gamma_{P,m}; \quad m = 1, 2 \tag{44}$$

and $\Gamma = -2 \operatorname{Im} \Sigma^r$ for each contribution. Note that while the radiative decay terms $\gamma_P$ in (43) and $\Gamma_{P,m}$ in (44) appear as they should in these expressions, they can be ignored relative to the other relaxation terms present here. In other words, in this application we can ignore all radiative coupling except to the pumping mode, and the latter is taken to second order. We have checked that for the present choice of parameters Eq.(43) constitutes an excellent approximation provided that the level populations $n_m$ that enter in (43) are calculated self consistently (using only the couplings $V^{(M)}$ and $V^{(N)}$ since the $V^{(P)}$ effect of the radiative coupling is taken only two second order and representing the effect of the energy transfer interaction $V^{(N)}$ by the approximation (36)). The following points can now be made:

(a) The absorption line shape, expressed by $I_{abs}(\omega_0)$, is dominated by the characteristic Lorentzian resonance shape centered about the molecular energy gap $\varepsilon_{21} = \varepsilon_2 - \varepsilon_1$. This is emphasized by considering the low bias case where $\mu_L$ and $\mu_R$ are both in the HOMO – LUMO gap, in which $\Gamma_1$ and $\Gamma_2$ are small relative to $\varepsilon_{21}$ as well as relative to the gaps between $\varepsilon_2$ and max($\mu_L$, $\mu_R$) and between $\varepsilon_1$ and min($\mu_L$, $\mu_R$). In this case we may take $f_L(E) = f_R(E) = 0$ and $f_L(E - \hbar\omega_0) = f_R(E - \hbar\omega_0) = 1$ (and consequently $n_2$=0 and $n_1$=1) in (43). This leads to the simple Lorentzian line shape

$$I_{abs}(\omega_0) = \frac{\left|V_0^{(P)}\right|^2}{\hbar} \frac{\Gamma}{(\varepsilon_2 - \omega_0 - \varepsilon_1)^2 + (\Gamma/2)^2} \frac{\Gamma_{M,1}\Gamma_{M,2}}{\Gamma_1\Gamma_2} \qquad (45)$$

with $\Gamma = \Gamma_1 + \Gamma_2$. Note that under the approximations that lead to (45), the term $\Gamma_{M,1}\Gamma_{M,2}/\Gamma_1\Gamma_2$ is nearly 1.

(b) The widths $\Gamma_1$ and $\Gamma_2$ in the denominators of Eq. (43) are the total level widths given by Eq. (44), where the radiative contribution $\Gamma_P = -2\,\text{Im}\,\Sigma_P^r$ has been disregarded compared with the other widths. The non-radiative width, $\Gamma_{NK,m}$, can be appreciable because of the small molecule-lead distance and should not be disregarded.

(c) The lineshape (43) shows an interesting dependence on the frequency ω₀ and on the bias potential Φ. Deviations from Lorentzian shape enter via the Fermi population functions, and reflect the partial population in the molecular resonances that interact with the metal electronic states. This effect depends on the imposed voltage through the voltage dependence of the electronic chemical potentials $\mu_L$ and $\mu_R$. We note in passing that an additional, trivial in the present context, voltage dependent effect is the Stark shift associated with the electric field in the biased cavity.

(d) Other effects of the junction environment on the spectroscopy enter via the dependence of the coupling $V_0^{(P)}$ on the metallic boundary conditions. This could be seen, in principle, in experiments that vary the inter-lead distance by stretching the molecular bridge.

## 4. Effect of electromagnetic driving on molecular conduction

As mentioned in section 1 there are many aspects of radiation field effect on conduction in molecular junctions. Theoretical treatments of transport in tunnel junctions in the presence of external oscillating fields were based on potential tunneling models[29], scattering based analysis of transport in mesoscopic junctions with oscillating parameters[30] or simple tight binding models with barrier or level energies and/or interstate coupling taken to oscillate with the frequency of the incident field.[9] To date, such effects were not observed in molecular junctions, though experimental effects of low frequency driving in larger mesoscopic junction have been reported.[10,31] Here we apply the model introduced in section 2 to discuss a different scenario where the radiation field is in resonance with the molecular optical transition.

Particularly interesting in this respect are molecules characterized by strong charge-transfer transitions that are reflected in the formation of a molecular excited state with a dipole far larger than that of the ground state dipole. For example, the dipole moment of DMEANS (4-Dimethylamino-4'-nitrostilbene) is 7D in the ground state and ~31D in the first excited singlet state,[32] For all-trans Retinal in Poly-methyl methacrylate films the dipole increase from ~6.6D to 19.8D upon excitation to the $^1B_u$ electronic state[33] and 40Å CdSe nanocrystals change their dipole from ~0 to ~32D upon excitation to their first excited state.[34] When such a species operates as a molecular wire connecting two metal leads with the direction of the optical charge transfer approximately parallel to the wire axis, optical pumping into the charge-transfer state creates an internal driving force for charge flow between the two leads. We may therefore expect that optical pumping that leads to the charge-transfer transition within the bridge can cause current flow in the absence of an imposed potential bias.

The implications of bridging two metal leads by such a molecule on details of the molecule-metal coupling are not known. Here we will make the reasonable assumption that a charge-transfer transition within the bridge is expressed in changing the relative coupling strengths of the molecular HOMO and LUMO to their metallic contacts. We thus investigate models in which $\Gamma_2 \neq \Gamma_1$ and $\Gamma_{ML,2} \neq \Gamma_{MR,2}$. The later inequality reflects the fact that the excited molecular state is dominated by atomic orbitals of larger amplitude on one side of the molecule than on the other and therefore with greater overlap with metal orbitals on that side.

The observable of interest is the induced electronic current. It is calculated from Eq.(7) by substituting $\Sigma_B^{<,>}$ by either $\Sigma_{ML}^{<,>}$ or $\Sigma_{MR}^{<,>}$,

$$\begin{aligned} I_{sd} &= \int_{-\infty}^{+\infty} \frac{dE}{2\pi\hbar} Tr\left[\Sigma_{ML}^{<}(E)G^{>}(E) - \Sigma_{ML}^{>}(E)G^{<}(E)\right] \\ &= -\int_{-\infty}^{+\infty} \frac{dE}{2\pi\hbar} Tr\left[\Sigma_{MR}^{<}(E)G^{>}(E) - \Sigma_{MR}^{>}(E)G^{<}(E)\right] \end{aligned} \tag{46}$$

The SEs $\Sigma_{MK}^{<,>}$ ($K=L,R$) are given by Eqs.(11). The GFs $G^{<,>}$ are obtained using the self-consistent procedure described in the section 3. The corresponding electronic current (Fig. 3) is $eI_{sd}$ where $e$ is the electronic charge.

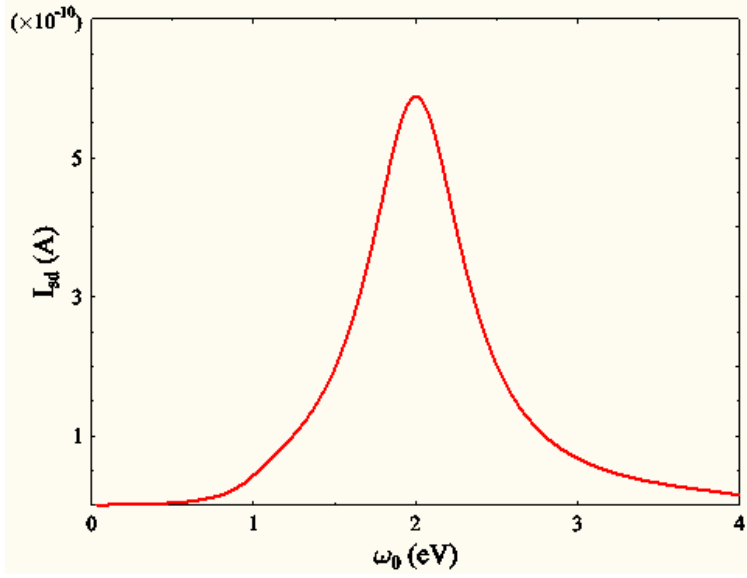

Fig.3 The photocurrent, Eq.(46), plotted against the incident light frequency in the absence of external potential bias. See text for parameters.

Figure 3 shows the resulting behavior – current induced by light without potential bias, obtained from Eq.(46) using the full self-consistent calculation described in Sect. 2. The parameters used in this calculation are $\Phi = 0$ (i.e. $\mu_L = \mu_R$), $\Gamma_{ML,1} = \Gamma_{MR,1} = 0.2\,eV$, $\Gamma_{ML,2} = 0.02\,eV$ and $\Gamma_{MR,2} = 0.3\,eV$. The other parameters are taken as in Fig. 2: $T = 300\,K$, $\varepsilon_2 - \varepsilon_1 = 2\,eV$, $\gamma_P = 10^{-6}\,eV$ $V_0^{(P)} = 10^{-3}\,eV$ and $B_{NL} = B_{NR} = 0.1\,eV$. As expected, steady state current flows through the junction in the presence of pumping. Naturally one gets a current peak at the frequency of the charge transfer transition, i.e. the HOMO-LUMO energy gap in our model.

The fact that photocurrent can occur in a molecular junction model with the postulated characteristics is a direct consequence of the fact that the charge transfer properties of the bridge lead to an internal driving force that would result in photovoltage in the corresponding open circuit. The critical question is whether such currents are of magnitudes that can be observable. The numbers taken above for the $\Gamma_{MK,m}$ parameters ($K=L,R$; $m=1,2$) are reasonable, and in any case we find that similar results are obtained when they are changed within a reasonable range. Also, the choice $B_{NL} = B_{NR} = 0.1 eV$ reflects an assumed lifetime of ~ 6 fs for an excited molecule at the metal surface to relax via the electron-hole pair mechanism – also a reasonable number. As indicated above, the results of Fig.3 were obtained using $V_0^{(P)} = \mu \mathcal{E} = 0.001 eV$, and where found to scale like $V_0^{(P)2}$ (i.e. like the radiation intensity) in our range of parameters. Here $\mu$ is the molecular transition dipole and $\mathcal{E}$ - the electric field associated with the electromagnetic radiation. If the charge-transfer transition dipole is taken as 1 Debye it would imply incident radiation intensity ($c|\mathcal{E}|^2/4\pi$ with $c$ – speed of light) ~ $10^8 \, watt/cm^2$ in vacuum. This number is of the order of magnitude of normal strong laser intensities used in spectroscopy, and it should be kept in mind that it could result from weaker incident fields due to local field enhancement that can take place in such geometries.[17,19,20] Another point of concern is the junction thermal stability under the proposed illumination. On the other hand, the current calculated with these parameters (see Fig. 3) is of order ~1nA, implying that radiation intensity lower by three orders of magnitude can still lead to observable currents. We conclude that photocurrent in single molecule junctions is a realistic possibility.

As in Section 3, we can gain insight on the predicted behavior by considering an approximation similar to that which yields Eq. (43). To this end we disregard $\Sigma_P^{<,>}$ in calculating $G^{<,>}$ from Eq.(12). However $\Sigma_{P0}^{<,>}$ is included, again to lowest order, in the calculation of $G^{<,>}$ in order to obtain the lowest order term in the effect of the radiation field on the current. This approximation yields (see Appendix C)

$$
\begin{aligned}
I_{sd} &= \int \frac{dE}{2\pi\hbar} \sum_{m=1,2} \Gamma_{ML,m} G^r_{mm}(E) \Gamma_{MR,m} G^a_{mm}(E) [f_L(E) - f_R(E)] \\
&+ |V_0^{(P)}|^2 \int \frac{dE}{2\pi\hbar} \frac{1}{(E-\varepsilon_2)^2 + (\Gamma_2/2)^2} \frac{1}{(E-\omega_0-\varepsilon_1)^2 + (\Gamma_1/2)^2} \\
&\times \{\Gamma_{ML,1}\Gamma_{MR,2} f_L(E-\omega_0)[1-f_R(E)] - \Gamma_{ML,2}\Gamma_{MR,1} f_R(E-\omega_0)[1-f_L(E)]\}
\end{aligned}
\quad (47)
$$

The first term on the right is the usual Landauer term that vanishes when the potential bias $\Phi$ is zero, i.e. $f_L = f_R$. The second term shows explicitly the effect of the pumping radiation to second order in the molecule-field coupling. In the absence of bias we set $f_L = f_R = f$ everywhere to get

$$
\begin{aligned}
I_{sd} &= |V_0^{(P)}|^2 \int \frac{dE}{2\pi\hbar} \frac{1}{(E-\varepsilon_2)^2 + (\Gamma_2/2)^2} \frac{1}{(E-\omega_0-\varepsilon_1)^2 + (\Gamma_1/2)^2} \\
&\times f(E-\omega_0)[1-f(E)] \{\Gamma_{ML,1}\Gamma_{MR,2} - \Gamma_{ML,2}\Gamma_{MR,1}\}
\end{aligned}
\quad (48)
$$

This expression shows explicitly how asymmetry in the HOMO and LUMO couplings to the metal electrodes leads to photocurrent in the present model. A further simplification is achieved when ω is not too far from its resonance value $\varepsilon_2 - \varepsilon_1$. In such case we can replace the term $f(E-\omega_0)[1-f(E)]$ by unity to get

$$
I_{sd} = \frac{|V_0^{(P)}|^2}{\hbar} \frac{\Gamma}{(\varepsilon_2 - \omega_0 - \varepsilon_1)^2 + (\Gamma/2)^2} \frac{\Gamma_{ML,1}\Gamma_{MR,2} - \Gamma_{ML,2}\Gamma_{MR,1}}{\Gamma_1 \Gamma_2}
\quad (49)
$$

As in Section 3, we have verified that for our choice of parameters the analytical result (48) provides an excellent approximation to the full self-consistent calculation.

The yield of this effect can be defined as the ratio

$$Y_c = \left(\frac{I_{sd}}{I_{abs}}\right)_{\Phi=0} \tag{50}$$

between the current induced in the unbiased junction, and the light absorbed by this junction. Using Eq. (49) for the former and Eq. (45) for the latter leads to

$$Y_c = \frac{\Gamma_{ML,1}\Gamma_{MR,2} - \Gamma_{ML,2}\Gamma_{MR,1}}{\Gamma_{M,1}\Gamma_{M,2}} \tag{51}$$

Again, this analytical approximation agrees with the full numerical calculation of this yield in the parameter region used.

To end this section we note that the situation discussed here, where each of the bridge HOMO and LUMO levels is coupled differently to the opposite leads can result in other interesting modes of behavior. For example, if we assume that the level position is pinned to the Fermi energy of the lead to which it is more strongly coupled it follows that the levels change their relative energies with the bias potential. As a result, the lineshape in Fig. 2 will shift under bias. More experimentally significant is the implication that changing bias under illumination with a fixed radiation frequency can take the molecule into and out of resonance with this radiation, leading to highly non-linear current voltage dependence including the possibility for negative differential resistance, see Fig.4.

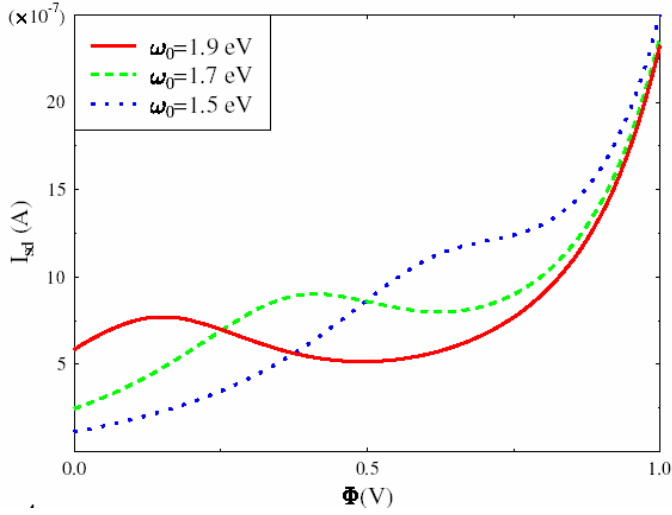

Fig.4 The source-drain current plotted against the voltage bias $\Phi$ obtained from Eq.(46) in the presence of light. The parameters used are $T = 300\,K$, $\varepsilon_{21} = 2\,eV$, Fermi level taken halfway between $\varepsilon_1$ and $\varepsilon_2$ in the absence of bias, $\Gamma_{ML,1} = \Gamma_{MR,2} = 0.2\,eV$, $\Gamma_{ML,2} = \Gamma_{MR,1} = 0.02\,eV$, $\gamma_P = 10^{-6}\,eV$, $B_{NL} = B_{NR} = 0.1\,eV$, $V_0^{(P)} = 0.02\,eV$. The bias $\Phi$ is assumed to shift the energies of the molecular orbitals according to

$$\varepsilon_m(\Phi) = \varepsilon_m(0) + \left(\Gamma_{ML,m} + \Gamma_{MR,m}\right)^{-1}\left[\left(\mu_L(\Phi) - \mu(0)\right)\Gamma_{ML,m} + \left(\mu_R(\Phi) - \mu(0)\right)\Gamma_{MR,m}\right],$$

$m = 1, 2$, where in the present calculation we took $\mu_L(\Phi) = \mu(0) + e\Phi$ and $\mu_R(\Phi) = \mu(0)$.

## 5. Fluorescence from current carrying molecular bridges

Light emission from STM junctions has been known for some time. Most studies of this effect have focused on emission from excited surface plasmons.[35][36] The process is pictured as resulting from the inelastic interaction of the tunneling electron with a surface plasmon, in which the latter is excited and later emits. This mechanism depends on geometrical parameters that determine the plasmon frequency, and on the electronic response properties of the leads that determine the lineshape and radiative yield of the plasmon emission. Alternatively, emission can originate within the molecular bridge of a molecular conduction junctions. [15,37] The mechanism for such emission could be similar to that pictured above, i.e. the time dependent potential of a tunneling electron causing electronic excitation of the molecule, However, assuming that the current in this case is actually carried by the molecule (i.e. through molecular orbitals) another picture emerges: in the steady-state current carrying situation the electronic distribution in the molecule is driven away from equilibrium and may be such that an electronic excited state is formed with a finite probability. In the language of single electron states this implies that a non-equilibrium electron hole distribution exists in the molecule, and if this distribution has electrons in otherwise unoccupied levels and holes in otherwise occupied ones, photon emission can take place. This mechanism is reminiscent of a light emitting diode operation, except that it now takes place in a single molecule. (&&&in ref for a discussion of a similar phenomenon of such effect in a metal nanodot see [16,38]) It is of interest to analyze the conditions under which such a mechanism can be operative, and to estimate the ensuing emission intensity and yield.

In our model the radiative fluxes can be obtained from Eq.(7) by using the self-energies $\Sigma_P(E,\omega)$, Eq. (30) and $\Sigma_P(E)$, Eq. (34), associated with the molecule-radiation field coupling. Note that in the absence of pumping all radiation field modes are treated on equal footing. The emission flux can be obtained as the net radiative flux obtained in the absence of pumping through either the lower state $|1\rangle$ or the higher state $|2\rangle$ with sign reversal. The frequency resolved spectrum is given by

$$I_{em}'(\omega) = \rho_P(\omega) \int_{-\infty}^{+\infty} \frac{dE}{2\pi\hbar} \left[ \Sigma_{P,11}^{<}(E,\omega)\, G_{11}^{>}(E) - \Sigma_{P,11}^{>}(E,\omega)\, G_{11}^{<}(E) \right]$$
$$= -\rho_P(\omega) \int_{-\infty}^{+\infty} \frac{dE}{2\pi\hbar} \left[ \Sigma_{P,22}^{<}(E,\omega)\, G_{22}^{>}(E) - \Sigma_{P,22}^{>}(E,\omega)\, G_{22}^{<}(E) \right]$$
(52)

and the overall emission intensity is the corresponding integral over all $\omega$,

$$I_{em}^{tot} = \int_0^{\infty} d\omega\, I_{em}'(\omega)$$
$$\int_{-\infty}^{+\infty} \frac{dE}{2\pi\hbar} \left[ \Sigma_{P,11}^{<}(E)\, G_{11}^{>}(E) - \Sigma_{P,11}^{>}(E)\, G_{11}^{<}(E) \right]$$
$$- \int_{-\infty}^{+\infty} \frac{dE}{2\pi\hbar} \left[ \Sigma_{P,22}^{<}(E)\, G_{22}^{>}(E) - \Sigma_{P,22}^{>}(E)\, G_{22}^{<}(E) \right]$$
(53)

The SEs and GFs needed in (52) and (53) are again obtained using the self-consistent procedure described in Section 3.

To reduce computational effort the computations done below were carried out with the radiation field taken to be at zero temperature. This would lead to an artifact at very low junction voltages since a finite temperature junction would emit radiation when coupled to a zero temperature radiation field even in the absence of voltage. To eliminate the artifact we shift the emission current calculated from Eq.(52) and (53) according to

$$I_{em}(\Phi) \to I_{em}(\Phi) - I_{em}(0)$$
(54)

This subtraction effectively corrects for absorption from the actual finite temperature radiation field.

Some results of this theory are shown in Fig. 5. Fig. 5a shows the integrated emission intensity (photons/s$^{-1}$), Eq. (53), plotted against the bias voltage. The parameters used are $T = 300K$, $\varepsilon_{21} = 2\ eV$, $\Gamma_{MK,m} = 0.1\ eV$ $(K = L, R;\ m = 1, 2)$, $\gamma_P = 10^{-6}\ eV$ and $B_{NL} = B_{NR} = 0.1 eV$. As before, $\mu(\Phi = 0) = E_F$ is taken in the middle of the HOMO-LUMO gap. Here the voltage bias is taken to shift the chemical potentials of the right and left electrodes symmetrically with respect to the (fixed) molecular orbital energies. This leads to onset of light emission at the threshold $e\Phi \simeq \hbar\omega_{21} = 2eV$. Fig. 5b shows similar results from a calculation that uses the same parameters except that the damping rates associated with energy transfer to electron-hole pairs in the leads is taken ten times larger, i.e. $B_{NL} = B_{NR} = 1 eV$. We see that the source-drain current is almost unaffected, however photon emission is substantially reduced. Fig. 5c compares the yields $I_{em}^{tot}/I_{sd}$ obtained in the two cases. The yield, of order ~$10^{-6}$ implies that one photon is emitted per ~$10^6$ electrons that traverse the junction.

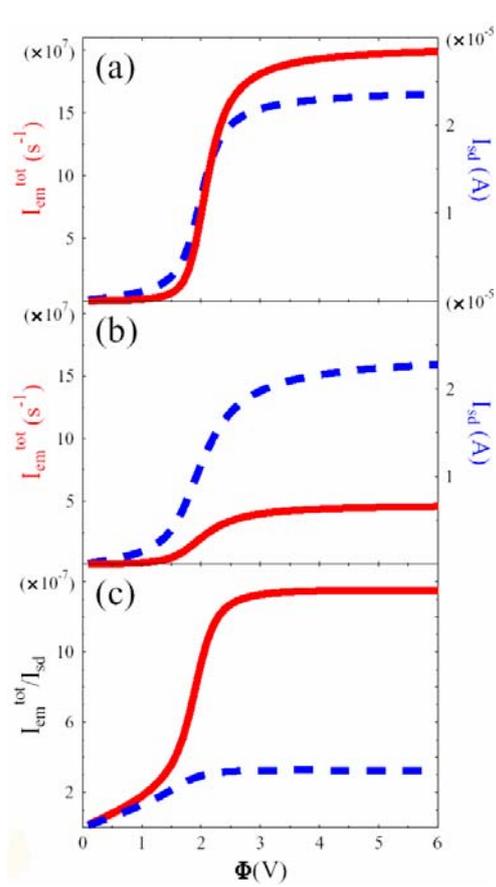

Fig. 5. (a) The integrated photon emission rate (full line; red) and the source-drain current (dashed line; blue) displayed as functions of the bias voltage using $T$=300K, $\varepsilon_{21}$=2 eV, $\Gamma_{MK,m}$ = 0.1 eV, ($K$=L,R; $m$=1,2), $\gamma_P = 10^{-6}\ eV$ and $B_{NL} = B_{NR} = 0.1 eV$. (b) Same as (a), except that $B_{NL}$

= $B_{NR}$ = 1eV. (c) The yield, $I_{em}^{tot}/I_{sd}$ plotted against the bias voltage for cases (a) – full line (red), and (b) – dashed line, blue.

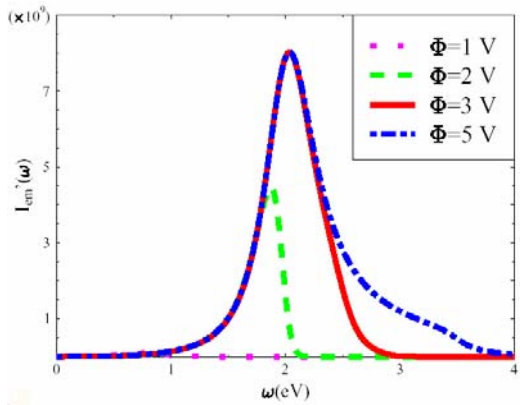

Fig. 6. Frequency resolved emission computed for the model of Fig. 1 using the parameters of Fig 5a, for different bias potentials.

Figure 6 shows, for the parameters of Fig. 5a, the frequency resolved emission for different bias potentials. Near the $\Phi = 2V$ threshold the higher frequency emission is cut off because of the partial electronic population in the metal-broadened HOMO and LUMO levels, a feature similar to what was seen in the absorption spectrum in Fig. 2.

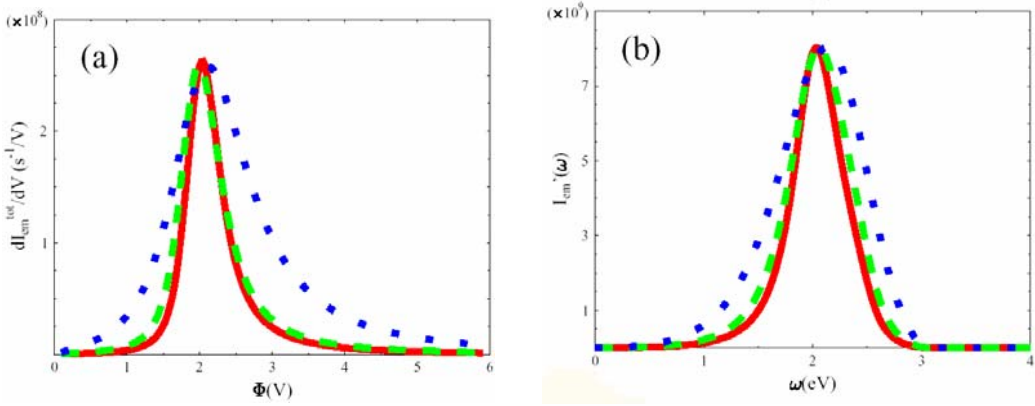

Fig. 7. The first derivative of the total emission intensity with respect to voltage (a) and the frequency resolved emission spectrum for a bias voltage $\Phi=3V$ (b). Full line (red) – parameters of Fig. 5a.. Dashed line (green) – same parameters except that $B_{NL}$ and $B_{NR}$ are taken larger by a factor 3. Dotted line (blue) same parameters as in fig. 5a except that $\Gamma_{MK,m}$ ($K=L,R$; $m=1,2$) are taken larger by a factor 3. All lines where scaled to the same height. In (a) this requires a multiplicative factor of 1.62 on the dotted line and 2.10 on the dashed line and in (b) the factors are 1.35 and 0.99 on the dotted and dashed lines, respectively.

Fig. 7 examines the effect of the damping processes associated with electron transfer ($\Gamma_{MK}$; $K=L,R$) and energy transfer ($B_{NK}$; $K=L,R$) on the derivative $dI_{em}/d\Phi$ plotted against $\Phi$ (Fig. 7a) and on the frequency resolved emission spectrum (Fig. 7b). These figures compare results obtained for the parameters used in Fig. 5a and for the cases either $\Gamma_M$ or $B_N$ are taken larger, with all lines normalized to the same peak height. Note that the width of the threshold region is more sensitive to the electron transfer rate $\Gamma_M$ than to the energy transfer rate $B_N$. This stems from the fact, implied by Eq. (37), that $B_N$ enters into the expression for the photoemission through terms like $B_N n_2$ and $B_N(1-n_1)$ where $n_1$ and $n_2$ are the occupations of levels 1 and 2 that that are smaller than 1 and for low bias satisfy $(1-n_1), n_2 \ll 1$.

Finally, Figure 8 shows results similar to Fig. 5, emission plotted against bias voltage, with parameters chosen to distinguish between a metal-molecule-metal contact as constructed by, e.g., break-junction or nanopore techniques, where the molecule is bound equally strongly to the two metal contacts, and an STM configuration where the molecule is bound strongly to the substrate and weakly to the tip. Within our model we assume that these differences are expressed in two ways. The first is the potential bias, $\Phi$, distribution in the junction. Defining the voltage division factor $\eta = \Phi_L/(\Phi_L + \Phi_R)$ where $\Phi_L$ and $\Phi_R$ ($\Phi_L + \Phi_R = \Phi$) are the magnitudes of the potential drops at the left and right molecule-lead contacts, we take $\eta = 0.5$ in the first case and $\eta \ll 1$ in the second (with the left lead representing the tip).[39] Secondly, this asymmetry is also reflected in the molecule-lead binding strength through our damping rate parameters $\Gamma_M$ and $B_N$. For definiteness we take $\Gamma_{ML,m} = (1-\eta)\Gamma_M; \Gamma_{MR,m} = \eta\Gamma_M; m = 1, 2$ and $B_{NL} = (1-\eta)B_N; B_{NR} = \eta B_N$. The full line in Fig. 8 reproduces the results of Fig. 5a (full line), the dashed line shows similar results for $\eta = 0.66$ and the dotted line was obtained using $\eta = 0.99$. The latter case, where the molecular energies are effectively pinned to the right lead and $\Gamma_{MR} \gg \Gamma_{ML}$; $B_{NR} \gg B_{NL}$, corresponds to an STM configuration with the tip considerably further from the molecule than the substrate.

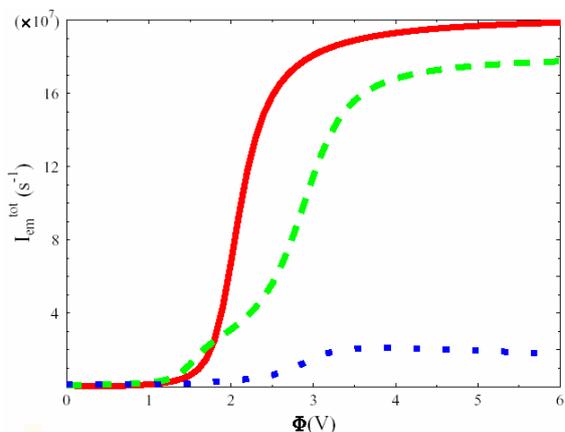

Fig. 8. Photon emission from junctions characterized by different voltage division factors (see text). Full line (red) η=0.5; dashed line (green) η=0.66; dotted line (blue) η=0.99.

An important feature in Fig 8 is that when the voltage bias is distributed unevenly between the two molecule-lead contacts, i.e. when the voltage division factor is different from 0.5, the threshold for photon emission moves to higher bias potentials, and when $\eta \to 1$ photon emission will not take place. Indeed, it is easy to see that starting from the Fermi energy in mid-gap and imposing a bias of this characteristic on the junction of Fig. 1 results in a situation where either the lower molecular level is always below the electrochemical potential of both leads, hence fully occupied or the upper molecular level is always above both electrochemical potentials, therefore fully empty – in either case no light emission can take place. Note that this does not exclude the possibility of light emission by other mechanisms, e.g. via plasmons excitation in the leads. It does suggest however that in STM junctions involving a molecular emitter, light emission from the molecule requires that part of the potential bias falls at the molecule-substrate interface. It is interesting to note that using a non-metalic conductor as a substrate or putting a spacer layer between the molecule and a substrate were suggested as ways to reduce energy losses into the substrate.[15] Within our model these are ways to reduce the parameter $B_N$. However, such measures also cause a potential drop at that interface – another important factor that enables light emission as discussed above.

As before, we can get analytical results for the emission current by taking the coupling to the radiation field to the lowest (second) order and using the approximate expressions (Appendix B) for the corresponding self-energies. This leads to (see Appendix C)

$$I_{em}'(\omega) = \frac{\gamma_P(\omega)}{\hbar} \int_{-\infty}^{+\infty} \frac{dE}{2\pi} \left\{ \frac{f_L(E+\omega)\Gamma_{ML,2} + f_R(E+\omega)\Gamma_{MR,2}}{(E+\omega-\varepsilon_2)^2 + (\Gamma_2/2)^2} \right.$$
$$\left. \times \frac{[1-f_L(E)]\Gamma_{ML,1} + [1-f_R(E)]\Gamma_{MR,1}}{(E-\varepsilon_1)^2 + (\Gamma_1/2)^2} \right\} \quad (55)$$

and

$$I_{em}^{tot} = \frac{\gamma_P(\varepsilon_{21})}{\hbar} n_2 (1-n_1) \quad (56)$$

where $\Gamma_m$, $m=1,2$ are given by Eq. (44) and $n_m$; $m=1,2$ are again the steady state populations in the molecular orbitals 1 and 2.

A more explicit expression for $I_{em}^{tot}$ is obtained in the limit where the potential bias is well above the threshold for light emission. In this case the chemical potential of one of the leads, for specificity let it be the left lead, is sufficiently above the LUMO energy while the chemical potential of the other is sufficiently below the HOMO energy.

In this case we show (Appendix C) that the partial contributions (see Eq. (C.3)) to the level populations $n_1$ and $n_2$ from the molecule-lead electron transfer interaction are

$$n_m^{MR} = 0; \quad n_m^{ML} = \Gamma_{ML,m}/\Gamma_m \quad ; \quad m = 1,2 \tag{57}$$

Using this with Eqs. (C.6) and (C.7) in Eq. (56) leads to

$$I_{em}^{tot} = \frac{\gamma_P}{\hbar} \frac{\Gamma_{ML,2}\Gamma_{MR,1}}{\Gamma_1 \Gamma_2} \tag{58}$$

Using expressions (58) and (C.11) we can get a simple expression for the yield of light emission in this limit. We get

$$Y_{em} \equiv \frac{I_{em}^{tot}}{I_{sd}} = \frac{\gamma_P}{\frac{\Gamma_{MR,2}}{\Gamma_{MR,1}}\Gamma_1 + \frac{\Gamma_{ML,1}}{\Gamma_{ML,2}}\Gamma_2 + B_N + \gamma_P} \tag{59}$$

It is interesting to note that the yield is enhanced in the asymmetric case where (for the bias direction indicated above) $\Gamma_{MR,2} < \Gamma_{MR,1}$ while $\Gamma_{ML,1} < \Gamma_{ML,2}$ — the inequalities required to promote light induced current in the unbiased junction as discussed in Section 4.

Again, we have verified that for our choice of parameters Eqs. (55)-(59) give results in close agreement with the numerical results displayed in Figs.5-7.

The results described above, that were obtained under rather conservative choices of damping parameters, indicate that this molecular mechanism for light emission from molecular conduction junctions yields measurable light intensities. An experimental and theoretical challenge is to find definitive ways to distinguish between this molecular mechanism and the plasmon dominated one. While no definitive statement can be made at this point, working with leads on which surface plasmons are strongly damped (so that no light is seen without molecules in the junction) seems to be a reasonable starting point.

In addition, recent advances in gated molecular junctions raise the possibility that one could distinguish between plasmon mediated and molecular light emission by the way they respond to an external gating potential.

## 6. Summary and conclusions

We have investigated several aspects of the interaction between a molecular conduction junction and the radiation field within a simple single-electron model that represents a molecular bridge by its highest occupied and lowest unoccupied levels, HOMO and LUMO, respectively. This model is the prototype of currently used models for molecular conduction, and its implications with regard to radiative effects stem from the fact that under potential bias, and in particular above the conduction threshold, the electronic distribution in the molecule is far from equilibrium. This non equilibrium situation is associated with three fluxes: First – electron flux between the source and drain leads, $I_{sd}$. Second, energy flux between the non-equilibriun molecular electronic distribution and the locally equilibrated electronic distributions in the metals. This is a non-radiative damping mechanism that couple electronic excitations in the molecule to electron-hole excitations in the metal. Finally, energy flux in the form of excited radiation field modes, i.e. light emission.

We have investigated several aspects of the interrelationships between these fluxes. First we have calculated the dependence of the optical absorption lineshape of the molecule as a function of the bias potential, secondly, we have studied condition under which current may be induced without potential bias using external light for driving a non-equilibrium steady state, and finally spontaneous light emission from a current carrying molecular junctions. These currents are obtained from a unified non-equilibrium Green function based formalism. While absorption is not an easily accessible observable in such junctions, we have concluded, using a range of reasonable parameter, that light driven electronic currents and current driven light emission are realistic possibilities. In addition to this statement about feasibility we have found several more interesting aspects of potential experimental significance:

(a) The use of molecular bridge that as isolated molecules are characterized by considerable difference between the electronic charge distributions in the ground and excited states, in particular molecules that show a large change in their electronic dipole moment upon electronic excitations, provides an interesting possibility for constructing junctions where a local electromotive force can be induced by light.

(b) While a natural threshold for strong electronic-optical correlation is the overlap between the molecular excitation spectrum and the bias potential, the possibility that the molecular electronic states and their energy depend on applied bias implies that the effects discussed above can behave in a strongly non-linear way. As seen in Fig. 4 this may appear as light induced negative differential resistance, but may also simply mean that the threshold for optical effects, when examined as a function of light frequency, will be bias dependent. Similarly, the emission lineshape discussed in Section 5 may be voltage dependent.

(c) The currents discussed above are characterized by their dependence on the frequency of the incident or emitted light and on the bias voltage. The widths of the corresponding spectra reflect the rates of both electron and energy transfer between the molecular bridge and its metallic environment. In turn, these widths affect the yield of the light induced current, or the current induced emission phenomena. These widths can be partially controlled by changing the lead material or inserting a suitable spacer between the molecule and the lead. Such manipulations also affect the way in which an imposed potential bias is distributed in the junction. In particular we have seen that the threshold of light emission from molecular junctions is a sensitive function of this distribution.

Due to competing relaxation processes, the yields of the light induced current and current induced emission that were the focus of our discussion are rather low. Nevertheless, as already known for the latter process, our estimates suggest that their observation is feasible. It should be kept in mind that in addition to relatively low yields, other obstacles do exist. Bringing light into a metal-molecule-metal junction is not a simple task though designs that make it possible do exist. Heating and the related issue of junction stability are also matters for concern. On the other hand, existing observations of photo-effects in molecular conduction junctions suggest that such experiments are indeed feasible. Correlating observations with predictions made in this paper should help the interpretation of future experiments in this direction.

**Acknowledgements.** This work was supported by the Israel Science Foundation the US-Israel Binational Science Foundation and by the NSF/NCN project through Northwestern University. The Authors thank Prof. Tobin Marks and Mark Ratner for helpful discussions, and Prof. Mark Ratner for hospitality and support during the time when this work was concluded.

## Appendix A: Derivation of general expressions for Green functions (GFs) and self energies (SEs)

Here we outline derivation of SEs used in the paper. We start from the Hamiltonian (1)-(6) with $\hat{H}_0$ being zero-order part and $\hat{V}$ representing perturbations. We seek expressions for Green functions such as $G_{ij}(\tau,\tau') = -i\langle T_c\, \hat{c}_i(\tau)\hat{c}_j^+(\tau')\rangle$ defined on the Keldysh contour with $\hat{T}_c$ being the time ordering operator (later times on left) on that contour. Expanding the contour evolution operator

$$\hat{S} = \exp\left[-i\int_c d\tau\, \hat{V}_I(\tau)\right] \quad (A.1)$$

in powers of $\hat{V}_I(\tau)$, the interaction representation of the coupling $\hat{V}$, and using this expansion in the GF of interest leads, after standard steps, to the Dyson equation for this GF

$$G_{ij}(\tau,\tau') = -i\langle T_c\, \hat{c}_i(\tau)\hat{c}_j^+(\tau')\rangle = G_{ij}^{(0)}(\tau,\tau') + \sum_{m,n}\int_c d\tau_1 \int_c d\tau_2\, G_{im}^{(0)}(\tau,\tau_1)\Sigma_{mn}(\tau_1,\tau_2) G_{nj}(\tau_2,\tau')$$

(A.2)

where $\Sigma$ is the corresponding SE. To find an explicit expression for $\Sigma$ we consider expansion up to second order in the interactions and work within the non-crossing approximation. The latter implies that different relaxation processes do not mix in $\Sigma$. This leads to the following expression for the GF

$$G_{ij}(\tau,\tau') \approx -i\left\langle T_c \hat{c}_i(\tau)\hat{c}_j^+(\tau')\left\{1 + \frac{(-i)^2}{2!}\int_c d\tau_1 \int_c d\tau_2 \right.\right.$$

$$\left[\sum_{K_1,K_2=L,R}\sum_{m_1,m_2=1,2}\sum_{\substack{k_1\in K_1 \\ k_2\in K_2}} \left(V^{(MK_1)}_{k_1,m_1}\hat{c}^+_{k_1}(\tau_1)\hat{c}_{m_1}(\tau_1) + h.c.\right)\left(V^{(MK_2)}_{k_2,m_2}\hat{c}^+_{k_2}(\tau_2)\hat{c}_{m_2}(\tau_2) + h.c.\right)\right.$$

$$+ \sum_{\alpha_1,\alpha_2}\left(V^{(P)}_{\alpha_1}\hat{a}_{\alpha}(\tau_1)\hat{c}^+_2(\tau_1)\hat{c}_1(\tau_1) + h.c.\right)\left(V^{(P)}_{\alpha_2}\hat{a}_{\alpha}(\tau_2)\hat{c}^+_2(\tau_2)\hat{c}_1(\tau_2) + h.c.\right) \qquad \text{(A.3)}$$

$$+ \sum_{K_1,K_2=L,R}\sum_{\substack{k_1\neq k_1'\in K_1 \\ k_2\neq k_2'\in K_2}} \left(V^{(NK_1)}_{k_1,k_1'}\hat{c}^+_{k_1}(\tau_1)\hat{c}_{k_1'}(\tau_1)\hat{c}^+_2(\tau_1)\hat{c}_1(\tau_1) + h.c.\right)$$

$$\left.\left.\left.\times \left(V^{(NK_2)}_{k_2,k_2'}\hat{c}^+_{k_2}(\tau_2)\hat{c}_{k_2'}(\tau_2)\hat{c}^+_2(\tau_2)\hat{c}_1(\tau_2) + h.c.\right)\right]\right\}\right\rangle$$

After opening parentheses and applying the Wick's theorem to the expression above, one gets the Dyson equation with SEs given by Eqs. (16)-(19).

**Appendix B: Approximate evaluation of the self energies**

In this work we encounter three types of self energy: $\Sigma_M$, the self energy associated with electron exchange between the molecule and the leads, whose retarded and advanced projections are simple imaginary constants in the wide band approximation, is given by Eqs. (9), (11) and (13). Here we derive approximate results for the self energies $\Sigma_P$ associated with the molecule-radiation field coupling, and $\Sigma_N$ associated with the interaction between molecular excitations and electron-hole pairs in the metal leads.

Consider first the self energies, $\Sigma_P^{>,<}$, Eq. (34), associated with the coupling to the radiation field. These functions contain, under the integral over ω, products of $\gamma_P(\omega)$, a relatively weak function of $\omega$, and $G^{>,<}_{mm}(E+\omega)$ ($m$=1,2) that according to Eq. (12) are products of a Lorentzian $G^r_{mm}(E\pm\omega)G^a_{mm}(E\pm\omega) = \left|G^r_{mm}(E\pm\omega)\right|^2$ and some function of energy $\Sigma^{>,<}(E)$. The Lorentzian $\left|G^r_{mm}(E\pm\omega)\right|^2$ is peaked about $E\pm\omega=\varepsilon_m$ ($m$=1,2), and if its width is small relative to range over which $\gamma_P(\omega)$ varies with ω, we can replace in Eq. (34a) $\Sigma^<_{P,11}(E)=(2\pi)^{-1}\int_0^\infty d\omega\,\gamma_P(\omega)\,G^<_{22}(E+\omega)$ by $(2\pi)^{-1}\gamma_P(\varepsilon_2-E)\int_E^\infty d\omega G^<_{22}(\omega)$ and $\Sigma^>_{P,22}(E)=(2\pi)^{-1}\int_0^\infty d\omega\,\gamma_P(\omega)\,G^>_{11}(E-\omega)$ by $(2\pi)^{-1}\gamma_P(E-\varepsilon_1)\int_0^\infty d\omega G^>_{11}(E-\omega)$.

Now the element $\Sigma^<_{P,11}(E)$ is important only near $E=\varepsilon_1$ because it enters in our flux calculation only in products with $G^<_{11}(E)$ (e.g., Eqs. (41), (53)) that peaks at this energy. Noting that $G^<_{22}(\omega)$ has a sharp peak near $\omega=\varepsilon_2$, we can write

$$\Sigma^<_{P,11}(\varepsilon_1)=(2\pi)^{-1}\gamma_P(\varepsilon_{21})\int_{\varepsilon_1}^\infty d\omega G^<_{22}(\omega) \quad =(2\pi)^{-1}\gamma_P(\varepsilon_{21})\int_{-\infty}^\infty d\omega G^<_{22}(\omega)=i\gamma_P(\varepsilon_{21})n_2$$

where $n_2=(2\pi i)^{-1}\int_{-\infty}^\infty dE\,G^<_{22}(E)$ is the population in level 2. Similarly, $\Sigma^>_{P,22}(E)$ is important near $E=\varepsilon_2$ and $\Sigma^>_{P,22}(\varepsilon_2)=(2\pi)^{-1}\gamma_P(\varepsilon_{21})\int_0^\infty d\omega\,G^>_{11}(\varepsilon_2-\omega)$ $=(2\pi)^{-1}\gamma_P(\varepsilon_{21})\int_{-\infty}^{\varepsilon_2}d\omega\,G^>_{11}(\omega)$ where again the upper integration limit can be set to ∞ yielding $\Sigma^>_{P,22}(\varepsilon_2)=(2\pi)^{-1}\gamma_P(\varepsilon_{21})\int_{-\infty}^\infty d\omega G^>_{11}(\omega)=-i\gamma_P(\varepsilon_{21})(1-n_1)$, where we have used for the population $n_1$ of level 1 the identity $1-n_1=-(2\pi i)^{-1}\int_{-\infty}^\infty dE\,G^>_{11}(E)$. We thus find that for our present application we can use

$$\Sigma^<_P(E)=i\gamma_P(\varepsilon_{21})\begin{bmatrix}n_2 & 0 \\ 0 & 0\end{bmatrix} \tag{B.1a}$$

$$\Sigma^>_P(E)=-i\gamma_P(\varepsilon_{21})\begin{bmatrix}0 & 0 \\ 0 & 1-n_1\end{bmatrix} \tag{B.1b}$$

$$\Gamma_P = i\left(\Sigma_P^> - \Sigma_P^<\right) = \gamma_P(\varepsilon_{21})\begin{pmatrix} n_2 & 0 \\ 0 & 1-n_1 \end{pmatrix} \tag{B.1c}$$

which are Eqs. (38).

Next consider the self energies $\Sigma_{NK}^{>,<}(E)$ ($K = L,R$), Eqs. (22) associated with energy transfer to electron-hole excitations in the metals. Specifically we will focus on $\Sigma_{NK}^<(E)$, Eq. (22a); the treatment of $\Sigma_{NK}^>(E)$ goes along similar lines. In the spirit of the wide band approximation we assume that $C_{NK}$, Eq. (24), is a constant. This implies, that at $T \to 0$ $B$ of Eq. (23) is essentially a step function

$$B_{NK}(\omega,\mu_K) = C_{NK}\omega\Theta(\omega) \tag{B.2}$$

Using this in (22a) we encounter integrals of the form $(2\pi)^{-1}\int_0^\infty d\omega\, \omega G_{22}^<(E+\omega)$ and $(2\pi)^{-1}\int_0^\infty d\omega\, \omega G_{11}^<(E+\omega)$ that can be approximated by $(2\pi)^{-1}(\varepsilon_2 - E)\int_E^\infty d\omega\, G_{22}^<(\omega)$ and $(2\pi)^{-1}(\varepsilon_1 - E)\int_E^\infty d\omega\, G_{11}^<(\omega)$, respectively. The first of these, i.e. the term involving $G_{22}^<$ appears in $\Sigma_{NK,11}^<$ and will appear in expression that peaks near $E = \varepsilon_1$. We can therefore replace this term by its value for $E = \varepsilon_1$ and use $\int_{\varepsilon_1}^\infty d\omega\, G_{22}^<(\omega) \simeq \int_{-\infty}^\infty d\omega\, G_{22}^<(\omega) = in_2$. The term involving $G_{11}^<$ appears in $\Sigma_{NK,22}^<$ and therefore its important contribution is near $E = \varepsilon_2$. However $\int_{\varepsilon_2}^\infty d\omega\, G_{11}^<(\omega) \simeq 0$ because even though $\int_{-\infty}^\infty d\omega\, G_{11}^<(\omega) = in_1$ most of the contribution comes from the neighborhood of level 1 that is out of the integral that has $\varepsilon_2$ as a lower bound. We thus conclude that under our assumptions

$$\Sigma_{NK}^< \simeq iB_{NK} \begin{pmatrix} n_2 & 0 \\ 0 & 0 \end{pmatrix} \quad \text{(B.3a)}$$

The same reasoning applied to $\Sigma_{NK}^>$ yields

$$\Sigma_{NK}^> \simeq -iB_{NK} \begin{pmatrix} 0 & 0 \\ 0 & 1-n_1 \end{pmatrix} \quad \text{(B.3b)}$$

so that the overall self energy associated with this process is

$$\Sigma_N^{>,<} = \Sigma_{NL}^{>,<} + \Sigma_{NR}^{>,<} \quad \text{(B.3c)}$$

and will have the same form as as the corresponding $\Sigma_{NK}^{>,<}$, with $B_N = B_{NL} + B_{NR}$ replacing $B_{NK}$. In Eqs. (B.3) $n_1$ and $n_2$ are again the populations of the corresponding levels. This result also implies that

$$\Gamma_{NK} = i\left(\Sigma_{NK}^> - \Sigma_{NK}^<\right) = B_{NK} \begin{pmatrix} n_2 & 0 \\ 0 & 1-n_1 \end{pmatrix} \quad \text{(B.4)}$$

**Appendix C: Simplified expressions for currents**

Here we outline the derivation of the approximate expressions (43), (47), (55) and (56) for the absorption, the light induced current and the emission flux. Our starting point is the Keldysh equation, Eq. (12), in which we use Eqs. (11), (29), (36) and (38) for the self energies and Eqs. (28), (27) and (16) for the retarded and advanced Green functions. We get

$$G_{11}^<(E) = \frac{if_L(E)\Gamma_{ML,1} + if_R(E)\Gamma_{MR,1} + i(B_N + \gamma_P)n_2 + 2\left|V_0^{(P)}\right|^2 G_{22}^<(E+\omega_0)}{(E-\varepsilon_1)^2 + (\Gamma_1/2)^2} \quad \text{(C.1a)}$$

$$G_{11}^{>}(E) = \frac{-i(1-f_L(E))\Gamma_{ML,1} - i(1-f_R(E))\Gamma_{MR,1} + |V_0^{(P)}|^2 G_{22}^{>}(E+\omega_0)}{(E-\varepsilon_1)^2 + (\Gamma_1/2)^2} \quad \text{(C.1b)}$$

$$G_{22}^{<}(E) = \frac{if_L(E)\Gamma_{ML,2} + if_R(E)\Gamma_{MR,2} + |V_0^{(P)}|^2 G_{11}^{<}(E-\omega_0)}{(E-\varepsilon_2)^2 + (\Gamma_2/2)^2} \quad \text{(C.1c)}$$

$$G_{22}^{>}(E) = \frac{-i(1-f_L(E))\Gamma_{ML,2} - i(1-f_R(E))\Gamma_{MR,2} - i(B_N+\gamma_P)(1-n_1) + 2|V_0^{(P)}|^2 G_{11}^{>}(E-\omega_0)}{(E-\varepsilon_2)^2 + (\Gamma_2/2)^2}$$
(C.1d)

The populations in the bridge levels are given by

$$n_m = -i \int_{-\infty}^{\infty} \frac{dE}{2\pi} G_{mm}^{<}(E) = 1 - i \int_{-\infty}^{+\infty} \frac{dE}{2\pi} G_{mm}^{>}(E); \quad m = 1,2 \quad \text{(C.2)}$$

It will be convenient to define also partial populations that correspond to the different process under consideration. We note that because $G^r$ is diagonal in the bridge subspace in our model, $G_{mm}^{<}(E) = |G_{mm}^{r}(E)|^2 \Sigma_{mm}^{<}(E)$ and that $\Sigma_{mm}^{<}$ contains additive contributions from these processes. We can therefore write

$$n_m = n_m^{ML} + n_m^{MR} + n_m^{NL} + n_m^{NR} + n_m^{P} \quad \text{(C.3)}$$

where

$$n_m^{\Theta} = -i \int_{-\infty}^{\infty} \frac{dE}{2\pi} |G_{mm}^{r}(E)|^2 \Sigma_{\Theta,mm}^{<}(E); \quad \Theta = ML, MR, NL, NR, P \quad \text{(C.4)}$$

For example, for the molecule-lead electron transfer interaction, this leads with our wide band approximation for the self energies to the familiar expression

$$n_m^{MK} = \int_{-\infty}^{+\infty} \frac{dE}{2\pi} \frac{\Gamma_{MK,m} f_K(E)}{(E-\varepsilon_m)^2 + (\Gamma_m/2)^2}; \quad K = L, R \quad \text{(C.5)}$$

Note that in the presence of a pumping mode there is in principle also a contribution $n_m^{P0}$ in (C.3), however this contribution is disregarded as we are considering the effect of this mode to the lowest (second) order and including such corrections amounts to taking higher order contributions into account. Using Eqs. (11), (36) and (38) then leads to

$$n_1 = n_1^{ML} + n_1^{MR} + \frac{B_N + \gamma_P}{\Gamma_1} n_2 \tag{C.6}$$

$$n_2 = n_2^{ML} + n_2^{MR} \tag{C.7}$$

where $\Gamma_{M,m} = \Gamma_{ML,m} + \Gamma_{MR,m}$.

The simplified expression for the electronic current $I_{sd}$ is now obtained by substituting Eqs. (11) and (C.1) into Eq. (46). This leads to

$$I_{sd} = I_{sd}^{(1)} + I_{sd}^{(2)} + I_{sd}^{(3)} \tag{C.8}$$

$$I_{sd}^{(1)} = \int_{-\infty}^{+\infty} \frac{dE}{2\pi\hbar} \sum_{m=1,2} \Gamma_{ML,m} G_{mm}^r(E) \Gamma_{MR,m} G_{mm}^a(E) [f_L(E) - f_R(E)] \tag{C.9a}$$

$$\begin{aligned} I_{sd}^{(2)} &= \frac{B_N + \gamma_P}{\hbar} \left[ n_2^{ML}(1 - n_1) - n_2 \left( \frac{\Gamma_{ML,1}}{\Gamma_1} - n_1^{ML} \right) \right] \\ &= \frac{B_N + \gamma_P}{\hbar} \left[ n_2^{ML} \left( \frac{\Gamma_{MR,1}}{\Gamma_1} - n_1^{MR} \right) - n_2^{MR} \left( \frac{\Gamma_{ML,1}}{\Gamma_1} - n_1^{ML} \right) \right] \end{aligned} \tag{C.9b}$$

$$I_{sd}^{(3)} = \frac{\left|V_0^{(P)}\right|^2}{\hbar} \int_{-\infty}^{+\infty} \frac{dE}{2\pi} \frac{1}{(E - \varepsilon_2)^2 + (\Gamma_2/2)^2} \frac{1}{(E - \omega_0 - \varepsilon_1)^2 + (\Gamma_1/2)^2} \times$$
$$\{ \Gamma_{ML,1} \Gamma_{MR,2} f_L(E - \omega_0)[1 - f_R(E)] - \Gamma_{ML,2} \Gamma_{MR,1} f_R(E - \omega_0)[1 - f_L(E)] + (B_N + \gamma_P)(1 - n_1 - n_2)$$
$$+ 2\Gamma_{ML,2} \Gamma_{MR,1} f_L(E)[1 - f_R(E - \omega_0)] - 2\Gamma_{ML,1} \Gamma_{MR,2} f_R(E)[1 - f_L(E - \omega_0)] \}$$
$$\tag{C.9c}$$

$I_{sd}^{(1)}$ is the flux due to the electron-transfer to leads interaction, and is the usual Landauer expression. $I_{sd}^{(2)}$ is the current associated with the radiative and non-radiative energy transfer interactions. $I_{sd}^{(3)}$ is the electronic current induced by the radiative pumping. Eqs. (C.9a) and (C.9c) may be simplified further. Using (following Eq.(C.4)) $n_m^{MK} = \int [dE/(2\pi)] G_{mm}^r G_{mm}^a \Gamma_{MK,m} f_K(E)$, (K=L, R), in (C.9a) leads to

$$I_{sd}^{(1)} = \frac{1}{\hbar} \sum_{m=1,2} \left[ \Gamma_{MR,m} n_m^{ML} - \Gamma_{ML,m} n_m^{MR} \right] \tag{C.10a}$$

Also, using in (C.9c) Eqs. (C.1) for the lesser and greater GFs, however disregarding terms higher than first order in $\left|V_0^{(P)}\right|^2$ (which means disregarding the terms that contain this factor in Eqs. (C.1) when used in (C.9c)) results in

$$I_{sd}^{(3)} = \frac{\left|V_0^{(P)}\right|^2}{\hbar} \int_{-\infty}^{+\infty} \frac{dE}{2\pi} \frac{1}{(E-\varepsilon_2)^2 + (\Gamma_2/2)^2} \frac{1}{(E-\omega_0-\varepsilon_1)^2 + (\Gamma_1/2)^2} \times$$
$$\{\Gamma_{ML,1}\Gamma_{MR,2} f_L(E-\omega_0)[1-f_R(E)] - \Gamma_{ML,2}\Gamma_{MR,1} f_R(E-\omega_0)[1-f_L(E)] + (B_N + \gamma_P)(1-n_1-n_2)$$
$$+ 2\Gamma_{ML,2}\Gamma_{MR,1} f_L(E)[1-f_R(E-\omega_0)] - 2\Gamma_{ML,1}\Gamma_{MR,2} f_R(E)[1-f_L(E-\omega_0)]\}$$
$$\tag{C.10b}$$

Note that in (C.9b) and (C.10b) the radiative width $\gamma_P$ can usually be disregarded relative to $B_N$ for a molecule near a metal surface.

These expressions become simpler in some limiting case. Under the experimental conditions addressed in Section 4, where we examine the possibility to induced current in an unbiased junction by resonance radiation, the molecular HOMO is below the leads Fermi energy and the LUMO is above it, so $n_1 \simeq 1$ and $n_2 \simeq 0$. This implies that $I_{sd}^{(2)} = 0$. Furthermore, in Eq. (C.10b) the main contribution to the integral comes from $E$ near $\varepsilon_2 \approx \varepsilon_1 + \omega_0$, and with no or small bias $[1 - f_K(E - \omega_0)] \simeq 0; (K = L, R)$ so the corresponding terms in (C.10b) can be disregarded. This leads to Eq. (47).

Consider now the experimental conditions of Section 5, where we discuss current induced light emission. In the limit where the bias is well above the threshold for emission the chemical potential of one lead, $\mu_L$ say, is sufficiently (compared to the level width) above the LUMO, while the other, $\mu_R$, is sufficiently below the HOMO. In this case $n_m^{NR} = 0$ and (using (C.5)) $n_m^{ML} = \Gamma_{ML,m}/\Gamma_M$ for $m = 1, 2$. Also, in this case $I_{sd}^{(3)} = 0$ because $V_0^{(P)} = 0$ (no pumping) and Eqs. (C.9b) and (C.10a) yield

$$I_{sd} = \frac{1}{\hbar} \sum_{m=1,2} \frac{\Gamma_{ML,m}\Gamma_{MR,m}}{\Gamma_m} + \frac{B_N + \gamma_P}{\hbar} \frac{\Gamma_{ML,2}\Gamma_{MR,1}}{\Gamma_1 \Gamma_2} \tag{C.11}$$

where again we can set $B_N + \gamma_P \simeq B_N$.

The calculations of the other currents proceed along similar lines. Using expressions (C.1) together with Eq. (29) in Eqs. (40) or (41), retaining only terms proportional to second order of the electron-pumping mode coupling leads to Eq. (43) for the absorption flux. Eqs. (55) and (56) for the differential and integrated emission fluxes are obtained from Eqs. (52) and (53), using Eqs. (30) and (C.1) for $\Sigma_P^{>,<}(E, \omega)$ in (52) and Eqs. (38) for $\Sigma_P^{>,<}(E)$ in (53). Again, we keep terms only up to second order in $V^{(P)}$ and note that no optical pumping exists in this case.

5421   M. Inoue and K. J. Ohtaka, Phys. Soc. Japan **52**, 3853 (1983).

22   A. Ottovaleitmannova and H. T. Tien, Progress in Surface Science **41** (4), 337 (1992).

23   A. Otto, Indian J. Phys. **77B**, 63 (2003).

24   Y. Meir and N. S. Wingreen, Phys. Rev. Lett. **68**, 2512 (1992); A. P. Jauho, N. S. Wingreen, and Y. Meir, Phys. Rev. B **50**, 5528–5544 (1994).

25   H. Haug and A.-P. Jauho, *Quantum Kinetics in Transport and Optics of Semiconductors*. (Springer, Berlin, 1996).

26   N. E. Bickers, Rev. Mod. Phys. **59**, 845 (1987).

27   D. C. Langreth, in *Linear and Nonlinear Transport in Solids*, edited by J. T. Devreese and D. E. v. Doren (Plenum Press, New York, 1976), pp. 3.

28   G. D. Mahan, *Many-particle physics*, 3 ed. (Plenum press, New York, 2000).

29   P. K. Tien and J. P. Gordon, Phys. Rev. **129**, 647–651 (1963).

30   O. Entin-Wohlman, A. Aharony, and Y. Levinson, Physical Review B **65** (19), 195411 (2002); M. Moskalets and M. Buttiker, Phys. Rev. B **69**, 205316 (2004).

31   A. H. Dayem and R. J. Martin, Phys. Rev. Letters **8**, 246–248 (1962); L. P. Kouwenhoven, A. T. Johnson, N. C. v. d. Vaart, C. J. P. M. Harmans, and C. T. Foxon, Phys. Rev. Lett. **69**, 1626–1629 (1991).

32   S. N. Smirnov and C. L. Braun, Rev. Sci. Instruments **69**, 2875 (1998).

33   M. Pondert and R. Mathies, J. Phys. Chem. **87**, 5090 (1983).

34   V. L. Colvin and A. P. Alivisatos, J. Chem. Phys. **97** (1), 730 (1992).

35   J. K. Gimzewski, B. Reihl, J. H. Coombs, and R. R. Schlittler, Zeitschrift fur Physik B **72**, 497 (1988); D. L. Abraham, A. Veider, C. Schonenberger, H. P.

56[39] W. D. Tian, S. Datta, S. H. Hong, R. Reifenberger, J. I. Henderson, and C. P. Kubiak, J. Chem. Phys. **109** (7), 2874 (1998).

# Figure captions

**Fig. 1.** A model for light induced effects in molecular conduction. The right ($R=|\{r\}\rangle$) and left ($L=|\{l\}\rangle$) manifolds represent the two metal leads characterized by electrochemical potentials $\mu_R$ and $\mu_L$ respectively. The molecule is represented by its highest occupied molecular orbital (HOMO), $|1\rangle$, and lowest unoccupied molecular orbital (LUMO), $|2\rangle$.

**Fig. 2** The absorption current (photons/s), Eq.(40) or (41) for the molecular model of Fig.1. The molecular electronic levels are assumed pinned to the right electrode, i.e. the bias shifts upward the electronic states of the left electrode. See text for parameters.

**Fig. 3** The photocurrent, Eq.(46), plotted against the incident light frequency in the absence of external potential bias. See text for parameters.

**Fig. 4** The source-drain current plotted against the voltage bias $\Phi$ obtained from Eq.(46) in the presence of light. The parameters used are $T = 300\,K$, $\varepsilon_{21} = 2\,eV$, Fermi level taken halfway between $\varepsilon_1$ and $\varepsilon_2$ in the absence of bias, $\Gamma_{ML,1} = \Gamma_{MR,2} = 0.2\,eV$, $\Gamma_{ML,2} = \Gamma_{MR,1} = 0.02\,eV$, $\gamma_P = 10^{-6}\,eV$, $B_{NL} = B_{NR} = 0.1\,eV$, $V_0^{(P)} = 0.02\,eV$. The bias $\Phi$ is assumed to shift the energies of the molecular orbitals according to

$$\varepsilon_m(\Phi) = \varepsilon_m(0) + \left(\Gamma_{ML,m} + \Gamma_{MR,m}\right)^{-1} \left[\left(\mu_L(\Phi) - \mu(0)\right)\Gamma_{ML,m} + \left(\mu_R(\Phi) - \mu(0)\right)\Gamma_{MR,m}\right],$$

$m = 1, 2$, where in the present calculation we took $\mu_L(\Phi) = \mu(0) + e\Phi$ and $\mu_R(\Phi) = \mu(0)$.

**Fig. 5.** (a) The integrated photon emission rate (full line; red) and the source-drain current (dashed line; blue) displayed as functions of the bias voltage using $T$=300K, $\varepsilon_{21}$=2 eV, $\Gamma_{MK,m}$ = 0.1 eV, ($K$=$L,R$; $m$=1,2), $\gamma_P = 10^{-6}\ eV$ and $B_{NL} = B_{NR} = 0.1\text{eV}$. (b) Same as (a), except that $B_{NL} = B_{NR}$ = 1eV. (c) The yield, $I_{em}^{tot} / I_{sd}$ plotted against the bias voltage for cases (a) – full line (red), and (b) – dashed line, blue.

**Fig. 6.** Frequency resolved emission computed for the model of Fig. 1 using the parameters of Fig 5a, for different bias potentials.

**Fig. 7.** The first derivative of the total emission intensity with respect to voltage (a) and the frequency resolved emission spectrum for a bias voltage $\Phi$=3V (b). Full line (red) – parameters of Fig. 5a.. Dashed line (green) – same parameters except that $B_{NL}$ and $B_{NR}$ are taken larger by a factor 3. Dotted line (blue) same parameters as in fig. 5a except that $\Gamma_{MK,m}$ ($K$=$L,R$; $m$=1,2) are taken larger by a factor 3. All lines where scaled to the same height. In (a) this requires a multiplicative factor of 1.62 on the dotted line and 2.10 on the dashed line and in (b) the factors are 1.35 and 0.99 on the dotted and dashed lines, respectively.

**Fig. 8.** Photon emission from junctions characterized by different voltage division factors (see text). Full line (red) η=0.5; dashed line (green) η=0.66; dotted line (blue) η=0.99.